\documentclass[a4paper,USenglish,cleveref, autoref, thm-restate]{lipics-v2021}

\usepackage[utf8]{inputenc} 
\usepackage[T1]{fontenc}    
\usepackage{hyperref}       
\usepackage{url}            
\usepackage{booktabs}       
\usepackage{amsfonts}       
\usepackage{nicefrac}       
\usepackage{microtype}      
\usepackage[dvipsnames]{xcolor}         
\usepackage{tikz}
\usetikzlibrary{positioning}

\hideLIPIcs
\AtBeginDocument{%
  }

\author{Denis Korotchenko}{ITMO University, St. Petersburg, Russia}{korotchenkods@gmail.com}{https://orcid.org/0009-0000-2068-7821}{}
\author{Vitaly Aksenov}{ITMO University, St. Petersburg, Russia \and \url{https://ctlab.itmo.ru/~vaksenov/}}{aksenov.vitaly@gmail.com}{https://orcid.org/0000-0001-9134-5490}{}
\authorrunning{D. Korotchenko and V. Aksenov}
\Copyright{Denis Korotchenko and Vitaly Aksenov}

\title{Semantic Lock: Synchronization Based on the Analysis of the Operation Conflict Graph}
\titlerunning{Semantic Lock: Sync. Based on the Analysis of the Operation Conflict Graph}



\EventEditors{XXX}
\EventNoEds{1}
\EventLongTitle{40th International Symposium on Distributed Computing (DISC 2026)}
\EventShortTitle{DISC 2026}
\EventAcronym{DISC}
\EventYear{2026}
\EventDate{November 9--13, 2026}
\EventLocation{Rome, Italy}
\EventLogo{}
\SeriesVolume{XX}
\ArticleNo{XX}

\begin{document}
\nolinenumbers


\maketitle

\begin{abstract}

This paper presents a new lock, SemanticLock, based on the conflict graph between operations. We can consider it a generalization of a read-write lock where conflicts exist between write operations and all other operations.

We demonstrate the effectiveness of our lock in two applications. In the first, we design a toy data structure: an array supporting point queries and different range queries. In the second, potentially of greater interest, we augment an existing concurrent data structure, ConcurrentHashMap, with additional long-running operations.

\vspace{0.5cm}

\noindent \textbf{AI Disclosure: We used GPT-5.5 to assist with improving the wording, clarity, and overall readability of the article. The authors verified the correctness and originality of all content, including references.}
\end{abstract}


\section{Introduction}
When developing applications for multicore systems, it is important to use efficient concurrent data structures. These data structures encapsulate thread synchronization logic, providing a user-friendly high-level interface that ensures correctness.

While commonly used implementations are highly optimized and reduce synchronization costs as much as possible, developers are often restricted by the operations they provide (e.g., just point queries). In practice, applications require custom complex operations (macro-operations) that span a large part of or even the entire data structure, such as computing an aggregate (e.g., summing elements), performing a global transformation, or taking a consistent snapshot. 

The fundamental problem addressed in this paper is how to simply and effectively incorporate such complex operations into existing lock-based data structures, while providing the standard correctness guarantee -- linearizability~\cite{Linearizability}.

One possible direction is to decrease the execution time of a long-running operation holding the lock, which spans different techniques: reusing blocked threads to accelerate the operation holding the lock~\cite{BrownCollaborative}; executing multiple operations in Flat Combining~\cite{hendler2010flat} under a single lock~\cite{Aksenov2019}; or applying partial persistence to speed up the execution or parallelization of specific operations~\cite{Verlib}.

Another direction, on which we focus, is to allow non-conflicting operations to run without excluding one another.
One of the most prominent approaches is to use transactional memory (TM)~\cite{shavit1995software}, including different techniques on top of it (e.g., range locks~\cite{Lomet}, range queries~\cite{rodriguez2025skip}). However, TM often incurs prohibitive metadata overhead and high abort rates, rendering it poorly suited for concurrent data structures with long-running operations. Conventional TM is oblivious to operation semantics and detects conflicts only during execution, thereby limiting potential performance gains.

A more direct use of operation semantics for concurrency control dates back to database compatibility matrices and semantic concurrency control~\cite{Badri}. In concurrent data structures, Transactional Boosting~\cite{Herlihy_Boosting} applies this idea by wrapping linearizable objects in software transactions and using commutativity-based locks. Multi-mode locks have also been studied in transactional settings~\cite{MML}. However, these approaches still require TM. Moreover, Transactional Boosting requires inverse operations, which may be difficult or impossible to provide for bulk operations. Retrofitting transactions into an existing non-transactional concurrent data structure can also be inefficient.

In contrast, SemanticLock, presented in this paper, exploits operation-level compatibility and targets shared-memory structures with a lightweight locking interface that requires neither TM, inverse operations, nor versioning. It acts as an access-control layer around ordinary operations, exposing fine-grained concurrency without TM overhead.

The first step toward SemanticLock is to use a read-write (RW) lock~\cite{RW}. In this case, we need to divide operations into two sets: those that can execute concurrently (readers) and those demanding exclusive access (writers). However, in some cases, RW locks are insufficiently fine-grained due to their bipartite nature. 

The next step leads toward Group Mutual Exclusion (GME) algorithms~\cite{GME}, which offer a more flexible abstraction: operations from the same group may access the resource concurrently, while operations from different groups exclude each other. However, GME still partitions operations into mutually exclusive groups, which may be insufficient in concurrent data structures, as we show by experiments.

Its generalization, Local Group Mutual Exclusion (LGME)~\cite{Luo2013LGME}, relaxes this global conflict assumption by introducing a conflict graph between groups: processes in the same group and processes in non-conflicting groups may execute simultaneously, whereas processes from conflicting groups must exclude each other. These abstractions are close in spirit to our goal because they exploit compatibility information rather than forcing exclusive access. However, LGME is designed as a distributed mutual exclusion problem based on group conflict relations and quorum/coterie constructions, whereas we focus on concurrent data structures.

Finally, complex data structures may expose arbitrary semantic dependencies among operation types: some operations can safely overlap with specific subsets of other operations, self-conflicts may differ from conflicts with other operations, and the desired mechanism is often a lightweight lock that can be added around existing operations. This motivates a solution that fully utilizes the conflict graph and can be easily integrated into a concurrent data structure when creating or expanding it.

In this paper, we present \texttt{SemanticLock}, apply it to a custom concurrent array, and use it to extend Java's \texttt{ConcurrentHashMap} with macro-operations. Our experiments show that the new structures outperform the versions with RW Lock and GME on workloads with compatibility patterns that these abstractions cannot express.

\section{Semantic Lock}
\subparagraph*{Conflict Graph Model.} \texttt{SemanticLock} is based on a concurrency control policy represented as an undirected graph $G=(V,E)$. Each method (or set of equally conflicting methods) of the target data structure is represented by a vertex $v\in V$. An edge $(u,v)\in E$ indicates a semantic conflict, meaning operations of types $u$ and $v$ cannot be executed concurrently. Note that $u$ and $v$ may be the same. In the current implementation, we require the user to provide this graph. Automatic graph generation via static analysis remains future work. We discuss the details of the implementation in Java.

\subparagraph*{Core Mechanism and Verification.} \texttt{SemanticLock} maintains, for each vertex $v \in V$, a counter $cnt[v]$, implemented as an \texttt{AtomicInteger} if $(v, v) \in E$ and as a \texttt{LongAdder} otherwise (see Appendix~\ref{app:atomic-problem} for a discussion of this choice). The value of $cnt[v]$ is the number of operations of type $v$ that are running and have not yet released the lock.

Upon invocation of an operation of type $u$, the thread repeatedly executes an optimistic \textit{acquisition phase} until it acquires permission to proceed. Let $N(u)=\{v\mid (u,v)\in E, v\ne u\}$ be the set of operation types that conflict with $u$, excluding $u$ itself. The acquisition phase consists of three steps:
\begin{enumerate}
    \item \textit{Precheck.} The thread reads all counters $cnt[v]$ for $v\in N(u)$. If some counter is positive, this acquisition attempt fails and the thread retries later. This step is only an optimization: it avoids modifying $cnt[u]$ when an already visible conflict exists.
    \item \textit{Reservation.} The thread reserves type $u$. If $(u,u)\in E$, the reservation is a compare-and-set on \texttt{AtomicInteger} $cnt[u]$ from $0$ to $1$; if it fails, the acquisition attempt fails. If $(u,u)\notin E$, the thread atomically increments \texttt{LongAdder} $cnt[u]$.
    \item \textit{Validation.} The thread reads all counters $cnt[v]$ for $v\in N(u)$ again. If all of them are zero, the acquisition succeeds and the operation may execute. Otherwise, the thread rolls back its reservation by decrementing $cnt[u]$ and retries the acquisition phase.
\end{enumerate}

After the operation completes, it releases \texttt{SemanticLock} by decrementing $cnt[u]$. Thus, an acquisition attempt may fail internally, but the data structure operation itself is not rejected: it simply retries until it obtains permission to run.

The optimistic verification protocol described above guarantees correctness, but not progress: several conflicting operations can reserve their types simultaneously and repeat the verification cycle infinitely. To avoid this, we extend \texttt{SemanticLock} with an extra global flag, which each thread tries to acquire (using \texttt{CAS(false, true)}) before decrementing if the \textit{validation} step fails. If the flag is acquired, the decrement is not performed and \textit{validation} step is repeated until success, while other conflicting threads fail at the \textit{precheck} step.

The correctness proof is deferred to Appendix~\ref{app:semanticlock-correctness}. In short, the \textit{reservation} and \textit{validation} steps ensure that two conflicting operation types cannot be admitted to run simultaneously.

However, a continuous stream of conflicting operations may cause some pending operation to starve. \texttt{SemanticLock} provides an optional fairness mode based on a request list. We explain the mechanism and prove that it provides starvation-freedom in Appendix~\ref{app:semanticlock-fairness}.

\section{Experiments}
\label{sec-experiments}
To evaluate \texttt{SemanticLock}, we implemented it in Java and integrated it with two data structures: a custom concurrent array with point and range operations and a \texttt{ConcurrentHashMap} extended with macro-operations. We compare it against Java's \texttt{ReentrantReadWriteLock} (\texttt{RW Lock}), two configurations of our GME implementation (\texttt{GME}), and a non-linearizable performance upper bound without synchronization (\texttt{No Lock}). We implemented GME ourselves because we did not find widely used Java implementation. We report throughput across several operation distributions. Finally, we used Lincheck~\cite{Lincheck} to test linearizability.

\subparagraph*{Testing Environment.} Benchmarks were conducted using a customized Synchrobench~\cite{Synchrobench} framework on a machine with an Intel Xeon Gold 5128 processor (16 physical cores, 32 logical threads via hyperthreading enabled) and 64 GB of RAM.

The framework initializes a data structure and spawns $T$ threads that repeatedly perform operations chosen according to a provided probability distribution. Each trial lasted 20 seconds: the first 10 seconds were used for warm-up, and throughput was measured during the remaining 10 seconds. We report averages over 10 independent runs after excluding outliers. In the experiments, we used SemanticLock with fairness mode disabled, because Synchrobench does not model starvation scenarios.

\begin{figure}[t]
	\begin{minipage}[t]{1.0\linewidth}
		\centering
        \begin{minipage}[t]{0.24\linewidth}
        \includegraphics[width=1.0\linewidth]{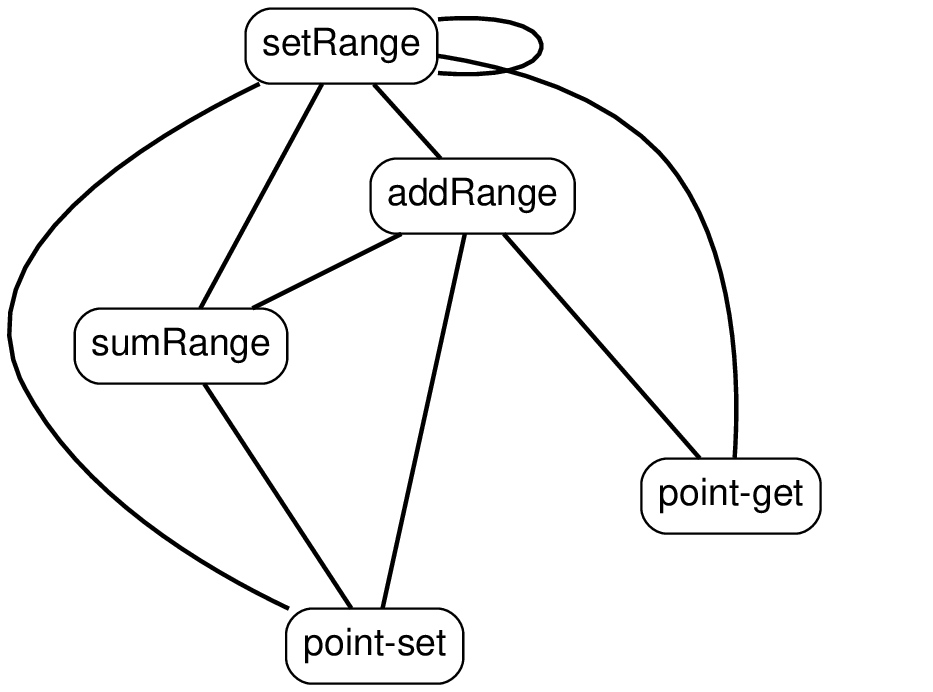}
        \small (a) SemanticLock \\conflict graph
        \end{minipage}
        \begin{minipage}[t]{0.24\linewidth}
        \includegraphics[width=1.0\linewidth]{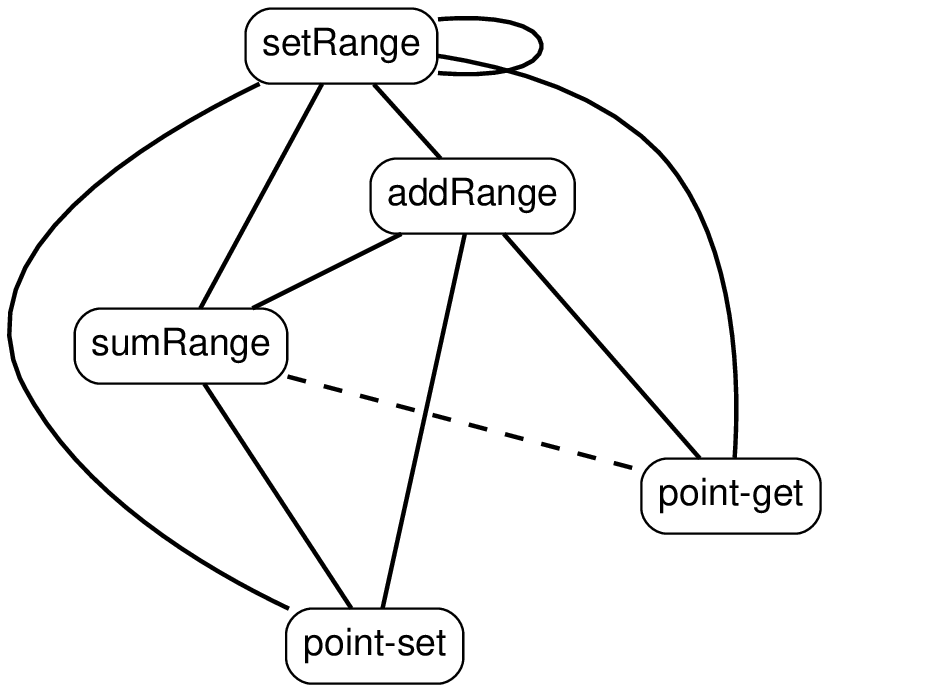}
        \small (b) Point-group GME \\ conflict graph
        \end{minipage}
        \begin{minipage}[t]{0.24\linewidth}
        \includegraphics[width=1.0\linewidth]{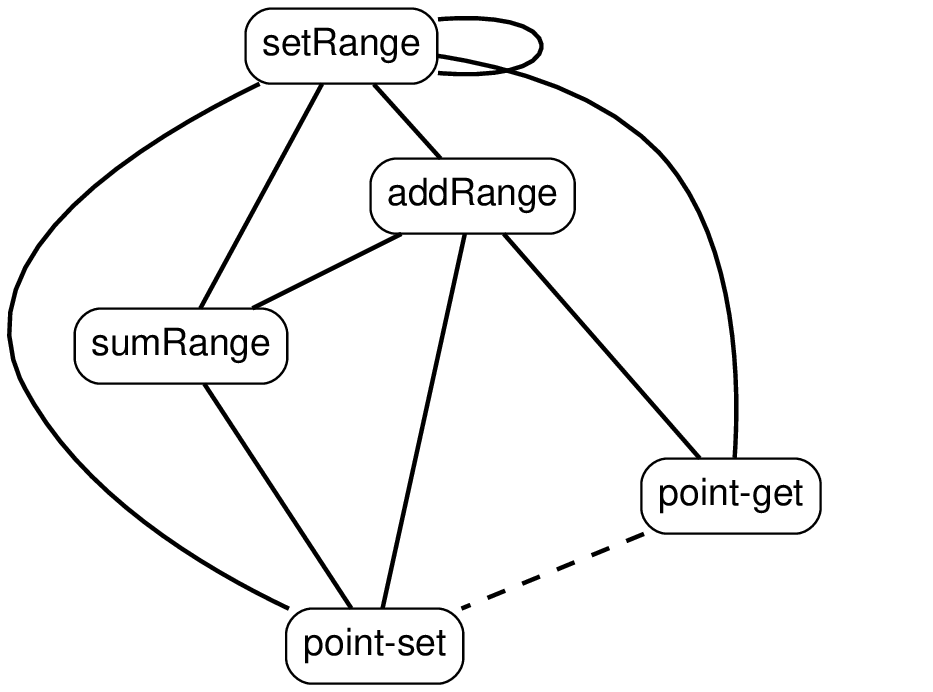}
        \small (c) Read-group GME \\ conflict graph
        \end{minipage}
        \begin{minipage}[t]{0.24\linewidth}
        \includegraphics[width=1.0\linewidth]{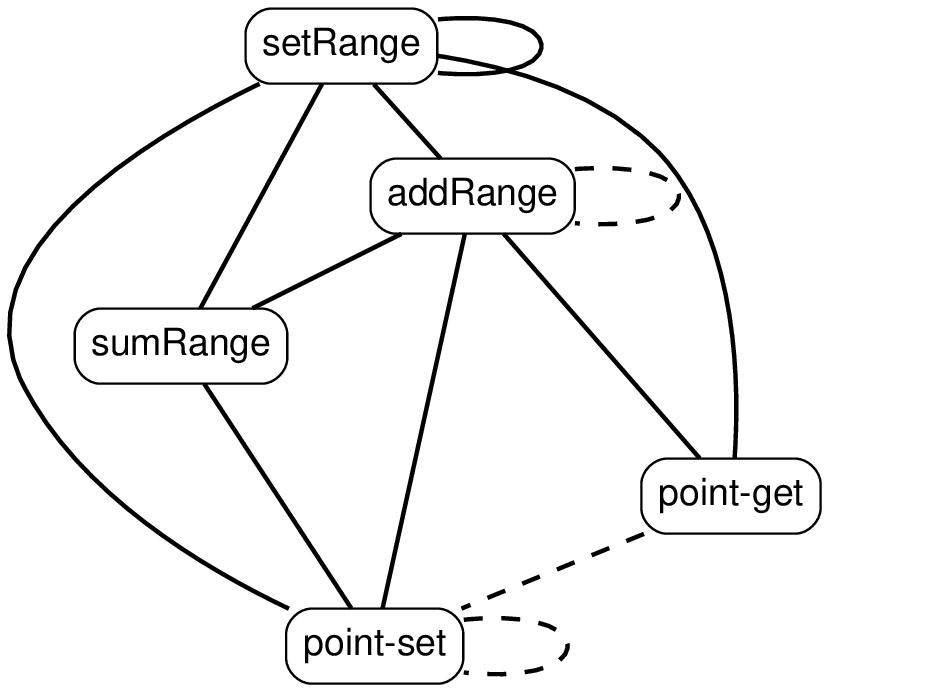}
        \small (d) Read-Write lock \\ conflict graph
        \end{minipage}
	\end{minipage}
	\caption{Semantics of the AtomicInteger array with range queries. Black edges present semantically non-compatible pairs and dashed edges present non-compatible pairs while using the corresponding algorithm. SemanticLock allows more pairs.}
	\label{fig:graph-ts}
\end{figure}

\subparagraph*{AtomicInteger Array with Range Queries.}
First, we evaluate a custom data structure: an array consisting of $N=2^{26}$ \texttt{AtomicInteger} elements. The structure supports five operations: \textsf{point-set(ind, x)}, \textsf{point-get(ind)}, \textsf{setRange(l, r, x)} (sets all integers in a range to $x$), \textsf{addRange(l, r, x)} (adds $x$ to all integers in a range), and \textsf{sumRange(l, r)} (returns the sum of integers in a range). For point operations, the index is chosen uniformly from $[0,N)$. For range operations, the range length $|r-l|$ is chosen uniformly from $[1,10^4]$, and the start index $l$ is chosen uniformly from $[0,N-|r-l|)$.

Figure~\ref{fig:graph-ts}(a) shows the compatibility relation encoded by \texttt{SemanticLock}; Figures~\ref{fig:graph-ts}(b) and (c) show the point-group and read-group GME partitions, and Figure~\ref{fig:graph-ts}(d) shows the RW-lock classification. GME must choose whether \textsf{point-get} shares a group with \textsf{point-set} or \textsf{sumRange}, since \textsf{point-set} conflicts with \textsf{sumRange}. In both variants, \textsf{addRange} forms a separate self-compatible group and \textsf{setRange} executes exclusively. An RW lock exposes only three compatible pairs, whereas \texttt{SemanticLock} provides full possible compatibility.

\begin{figure}[t]
	\begin{minipage}[t]{1.0\linewidth}
		\centering
		\begin{minipage}[t]{0.32\linewidth}
			\centering
			\includegraphics[width=1.0\linewidth]
            {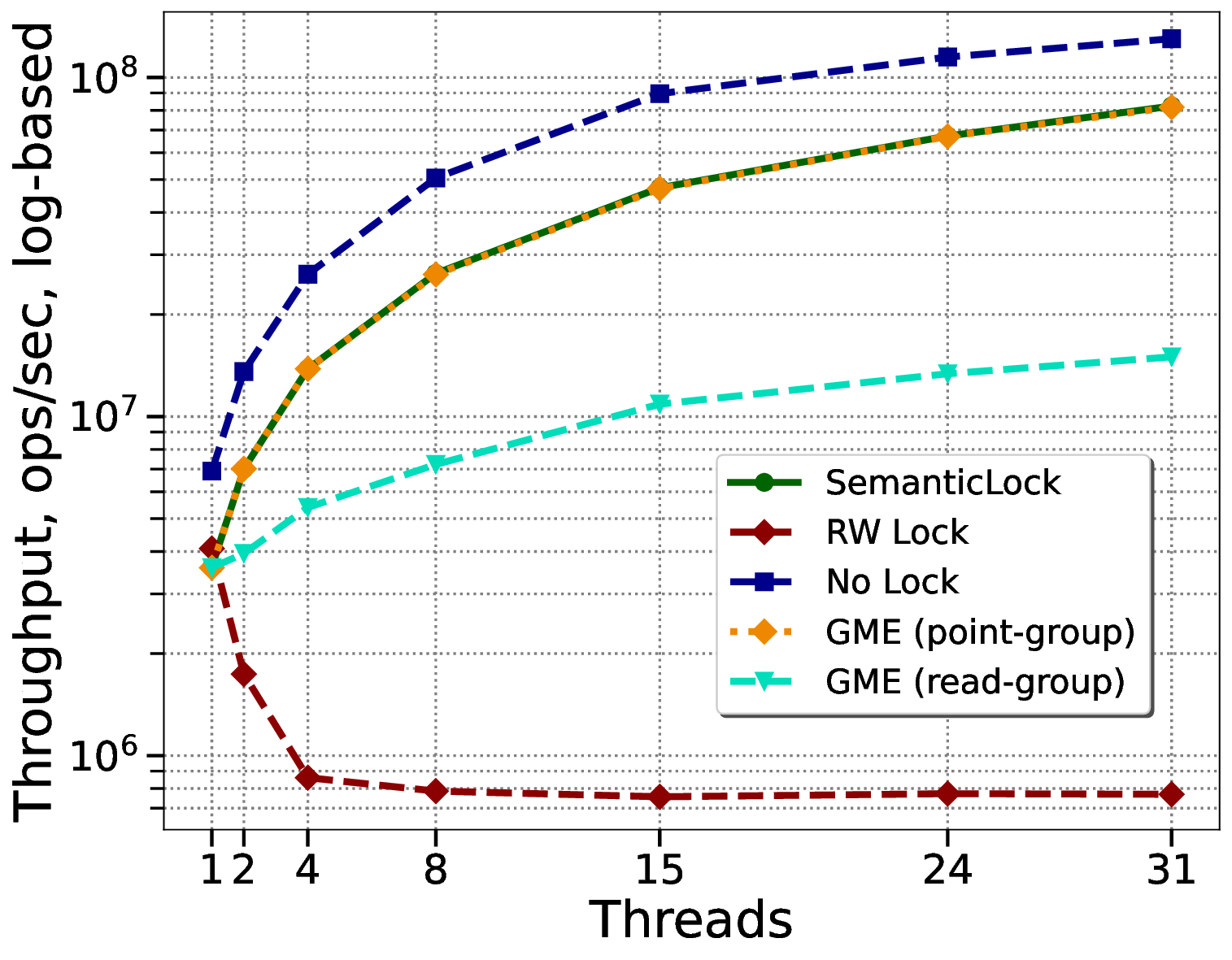}
			\small (a) $50\%$ set, $50\%$ get
		\end{minipage}
		\begin{minipage}[t]{0.32\linewidth}
			\centering
			\includegraphics[width=1.0\linewidth]
            {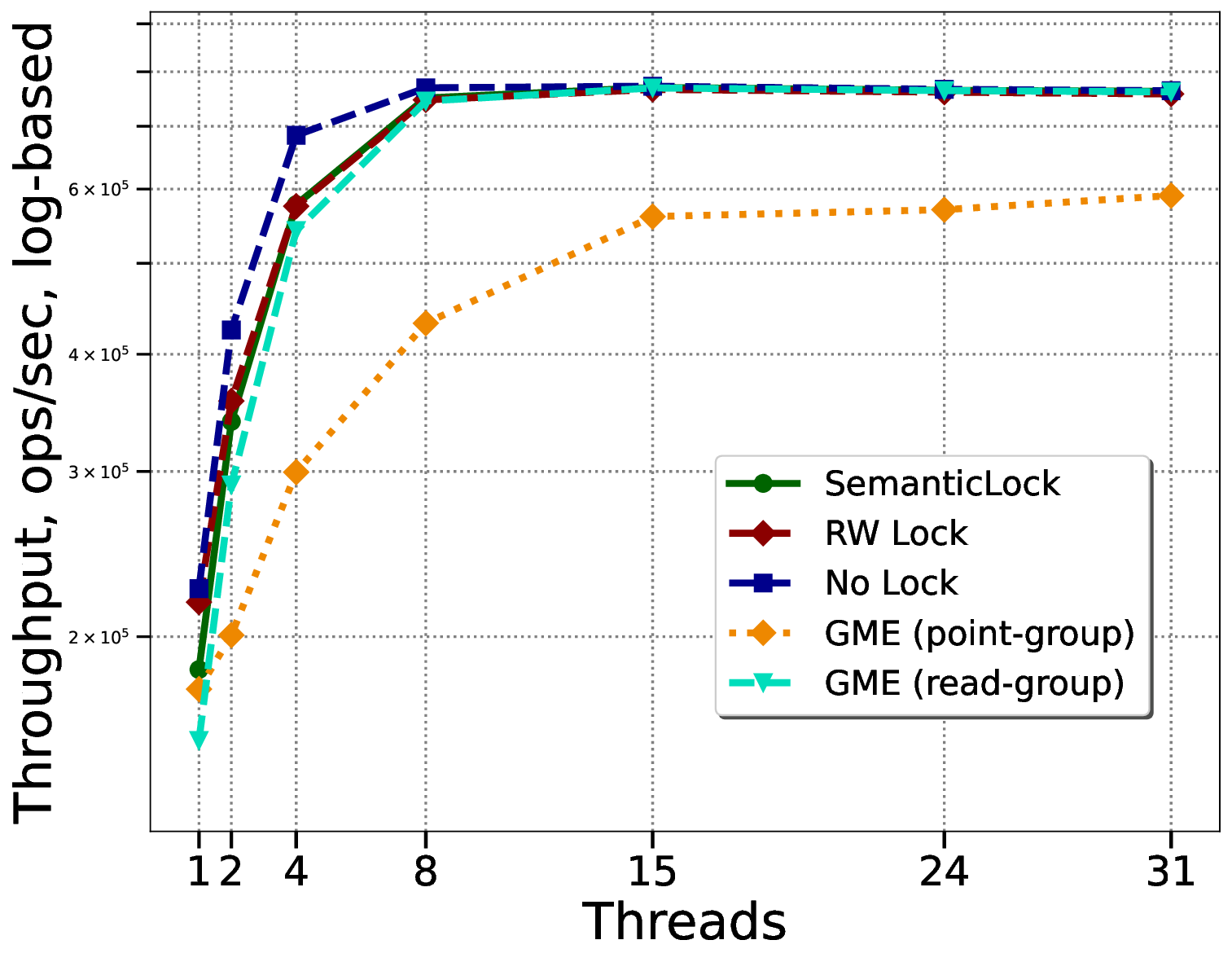}
			\small (b) $50\%$ get, $50\%$ sumRange
		\end{minipage}
        \begin{minipage}[t]{0.32\linewidth}
			\centering
			\includegraphics[width=1.0\linewidth]
            {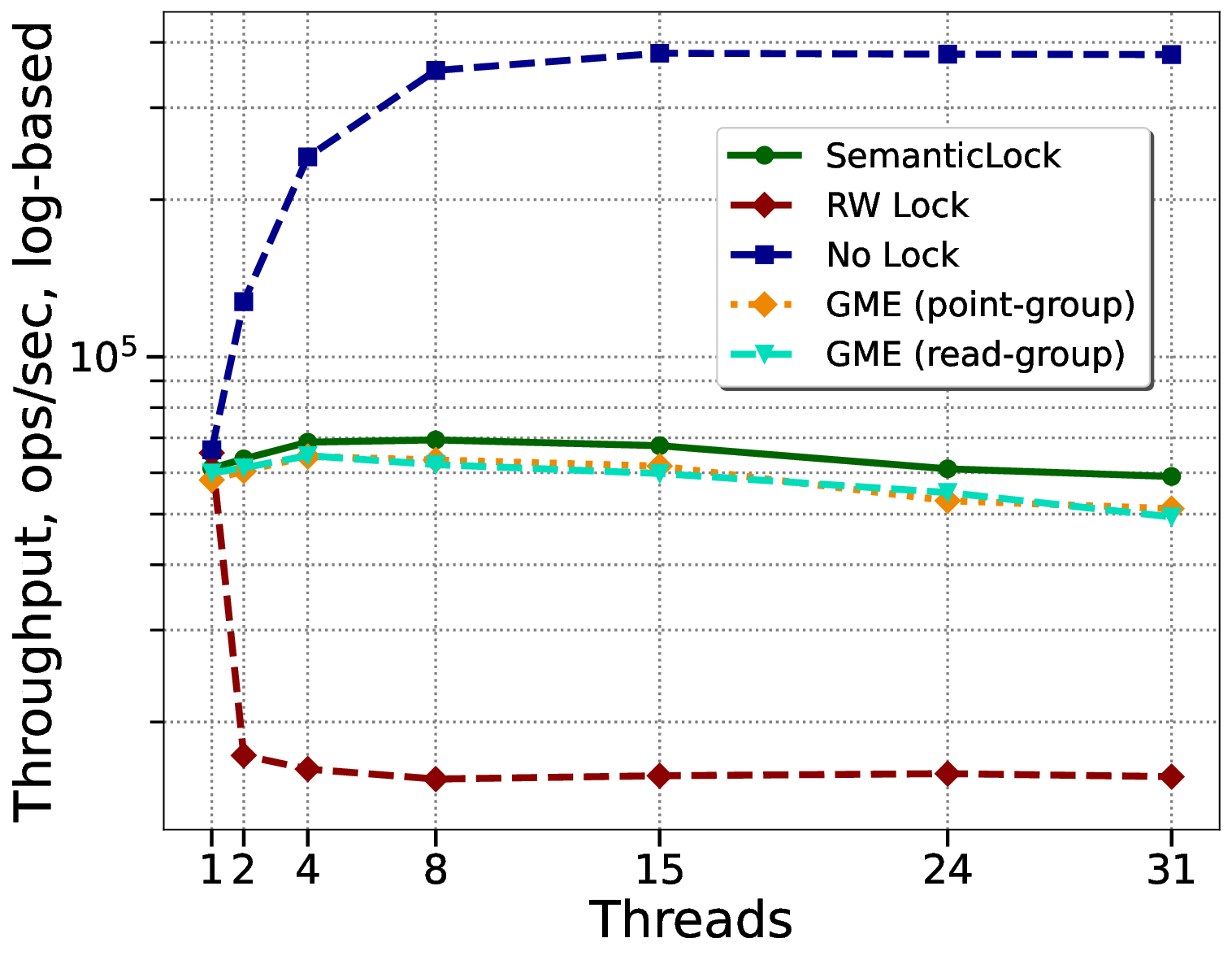}
			\small (c) $20\%$ set, $20\%$ get, \\$20\%$ sumRange,\\$20\%$ setRange, $20\%$ addRange
		\end{minipage}
	\end{minipage}

	\begin{minipage}[t]{1.0\linewidth}
		\centering
        \begin{minipage}[t]{0.32\linewidth}
			\centering
			\includegraphics[width=1.0\linewidth]
            {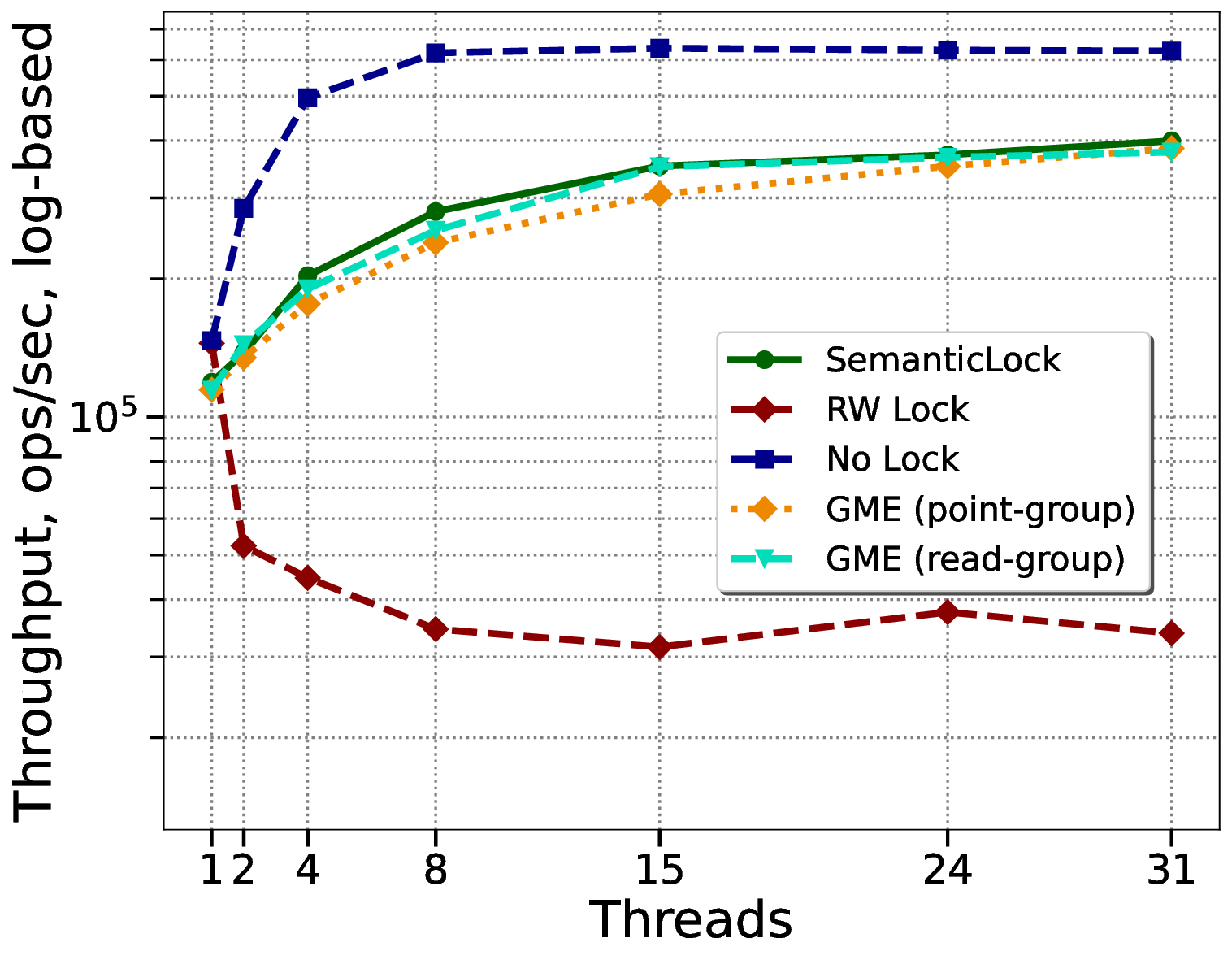}
			\small (d) $40\%$ get, $40\%$ sumRange, \\
             $10\%$ set, $10\%$ addRange
		\end{minipage}
		\begin{minipage}[t]{0.32\linewidth}
			\centering
			\includegraphics[width=1.0\linewidth]
            {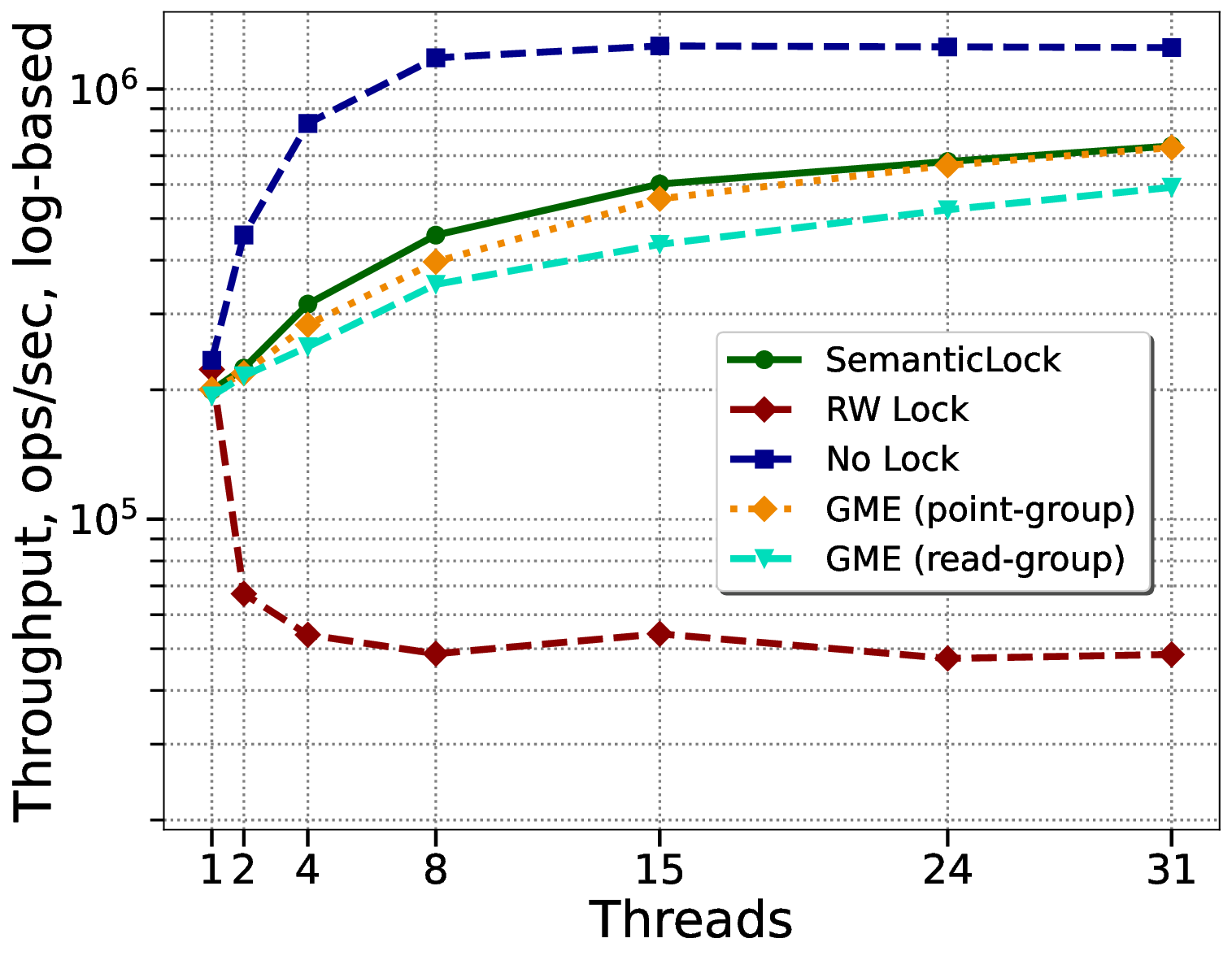}
			\small (e) $40\%$ get, $10\%$ sumRange, \\
             $40\%$ set, $10\%$ addRange
		\end{minipage}
        \begin{minipage}[t]{0.32\linewidth}
			\centering
			\includegraphics[width=1.0\linewidth]
            {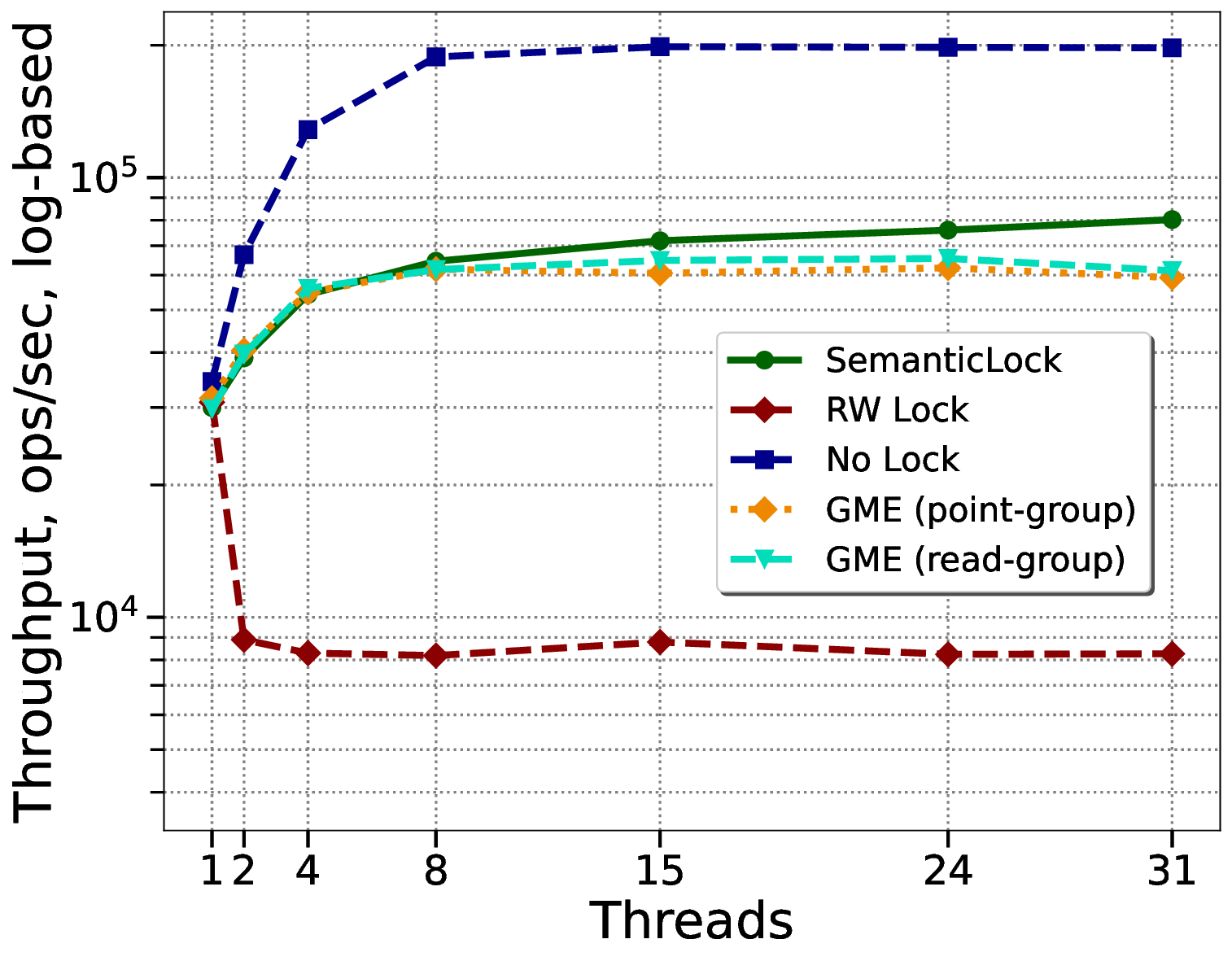}
			\small (f) $5\%$ sumRange,\\$5\%$ setRange, $90\%$ addRange
		\end{minipage}
	\end{minipage}
	\caption{Throughput of operations on AtomicInteger array under varying operation distributions.}
	\label{fig:array}
\end{figure}

Figure~\ref{fig:array} reports six representative workloads. In the point-operation and read-only workloads, shown in Figures~\ref{fig:array}(a) and~(b), SemanticLock incurs only verification and counter-update overhead while preserving scalability and achieving throughput close to that of \texttt{No Lock}.

For mixed workloads, the observed performance depends on the fraction of operation pairs that are compatible according to the conflict graph. In workloads dominated by concurrently admissible operations, such as point operations, range sums, and self-compatible range additions (Figures~\ref{fig:array}(d), (e) and~(f)), SemanticLock provides good scalability. In the workload with a larger fraction of conflicting operations (Figure~\ref{fig:array}(c)), scalability is more limited, but SemanticLock still outperforms RW Lock and both GME variants.

\subparagraph*{Extended ConcurrentHashMap.} Next, we extend Java's \texttt{ConcurrentHashMap} with two macro-operations: \textsf{keysSnapshot()}, which returns a snapshot of all keys, and \textsf{mapValues(lambda)}, which applies a function \textsf{lambda} to every value. In our experiments, \textsf{mapValues} applies \(\textsf{cap(value, limit)}=\min(\textsf{value}, \textsf{limit})\), where the limit may vary between invocations. Workloads also include point reads (\textsf{get}) and point updates (\textsf{put}/\textsf{remove}). Point-operation keys are chosen uniformly from $[0,2^{20})$; each key is initially present with probability $0.5$.

\begin{figure}[tb]
	\begin{minipage}[t]{1.0\linewidth}
        \centering
        \begin{minipage}[t]{0.24\linewidth}
            \centering
			\includegraphics[width=1.0\linewidth]{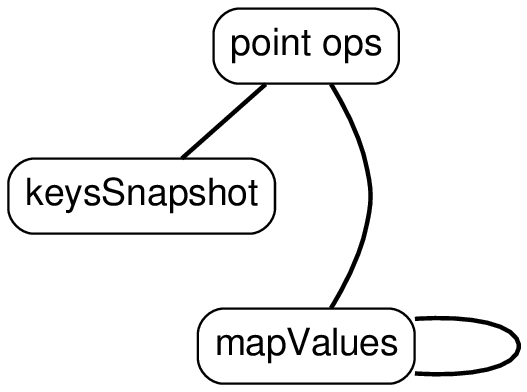}
            \small (a) \texttt{SemanticLock-3} conflict graph
        \end{minipage}
        \begin{minipage}[t]{0.24\linewidth}
            \centering
			\includegraphics[width=1.0\linewidth]{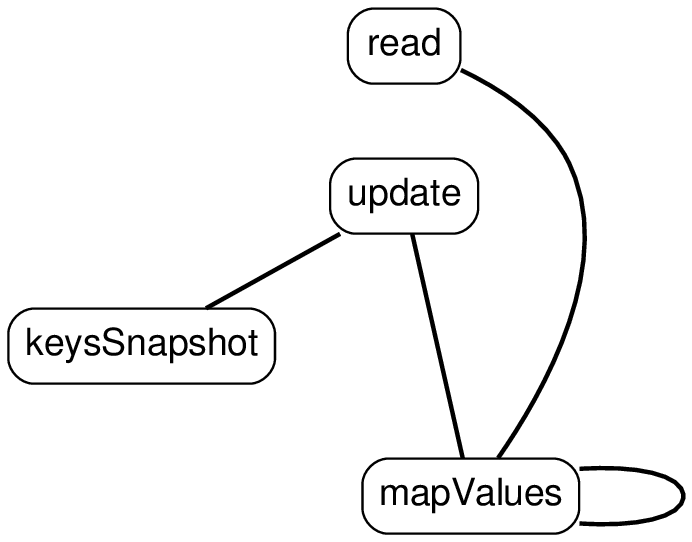}
            \small (b) \texttt{SemanticLock-4} conflict graph
        \end{minipage}
        \begin{minipage}[t]{0.24\linewidth}
            \centering
			\includegraphics[width=1.0\linewidth]{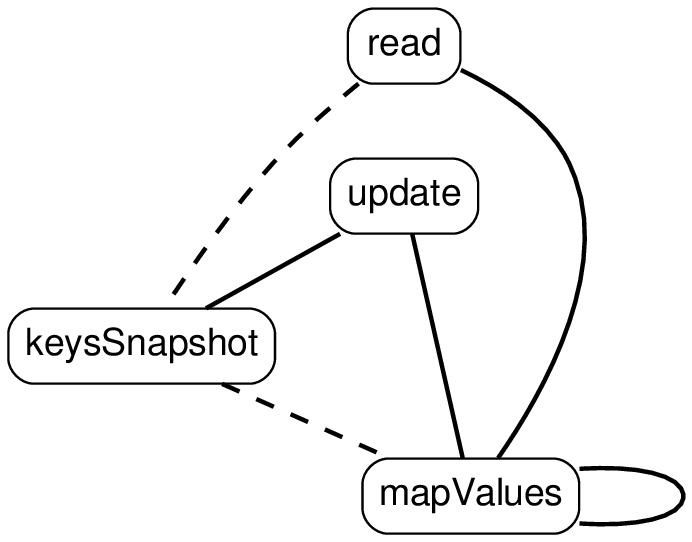}
            \small (c) Point-group GME  \\ conflict graph
        \end{minipage}
        \begin{minipage}[t]{0.24\linewidth}
            \centering
			\includegraphics[width=1.0\linewidth]{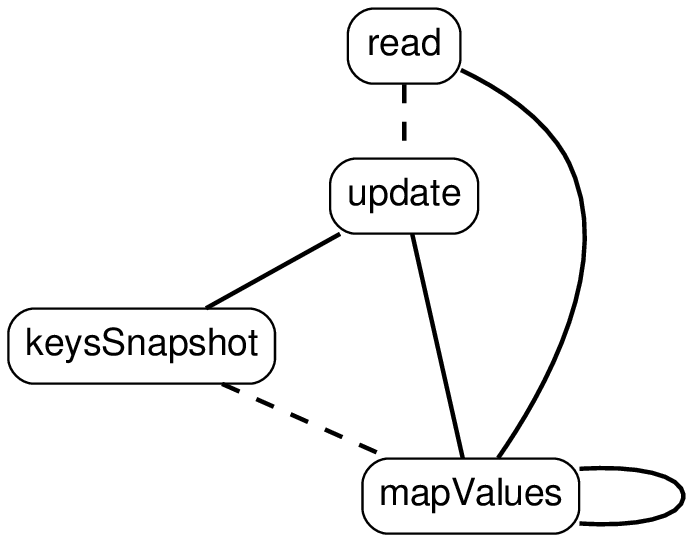}
            \small (d) Read-group GME \\ conflict graph
        \end{minipage}
	\end{minipage}
	\caption{Semantics of the extended ConcurrentHashMap. Black edges present semantically non-compatible pairs and dashed edges present non-compatible pairs while using the corresponding algorithm. SemanticLock allows more pairs.}
	\label{fig:graph}
\end{figure}

We evaluate two \texttt{SemanticLock} configurations: \texttt{SemanticLock-3}, whose coarse graph groups all point operations in one vertex (Figure~\ref{fig:graph}(a)), and \texttt{SemanticLock-4}, whose fine-grained graph separates \textsf{get} from \textsf{put}/\textsf{remove} (Figure~\ref{fig:graph}(b)). We compare them with \texttt{No Lock}, \texttt{RW Lock} (all point operations are readers, whereas both macro-operations are writers), and two GME configurations. In \texttt{GME (point-group)}, all point operations share one group; in \texttt{GME (read-group)}, \textsf{get} and \textsf{keysSnapshot} share one group. Figures~\ref{fig:graph}(c) and (d) illustrate the required choice: point updates conflict with \textsf{keysSnapshot} and therefore cannot share its group. An RW lock faces the same binary choice, whereas \texttt{SemanticLock} encodes all safe overlaps.

\begin{figure}[tb]
	\begin{minipage}[t]{1.0\linewidth}
		\centering
		\begin{minipage}[t]{0.32\linewidth}
			\centering
			\includegraphics[width=1.0\linewidth]
            {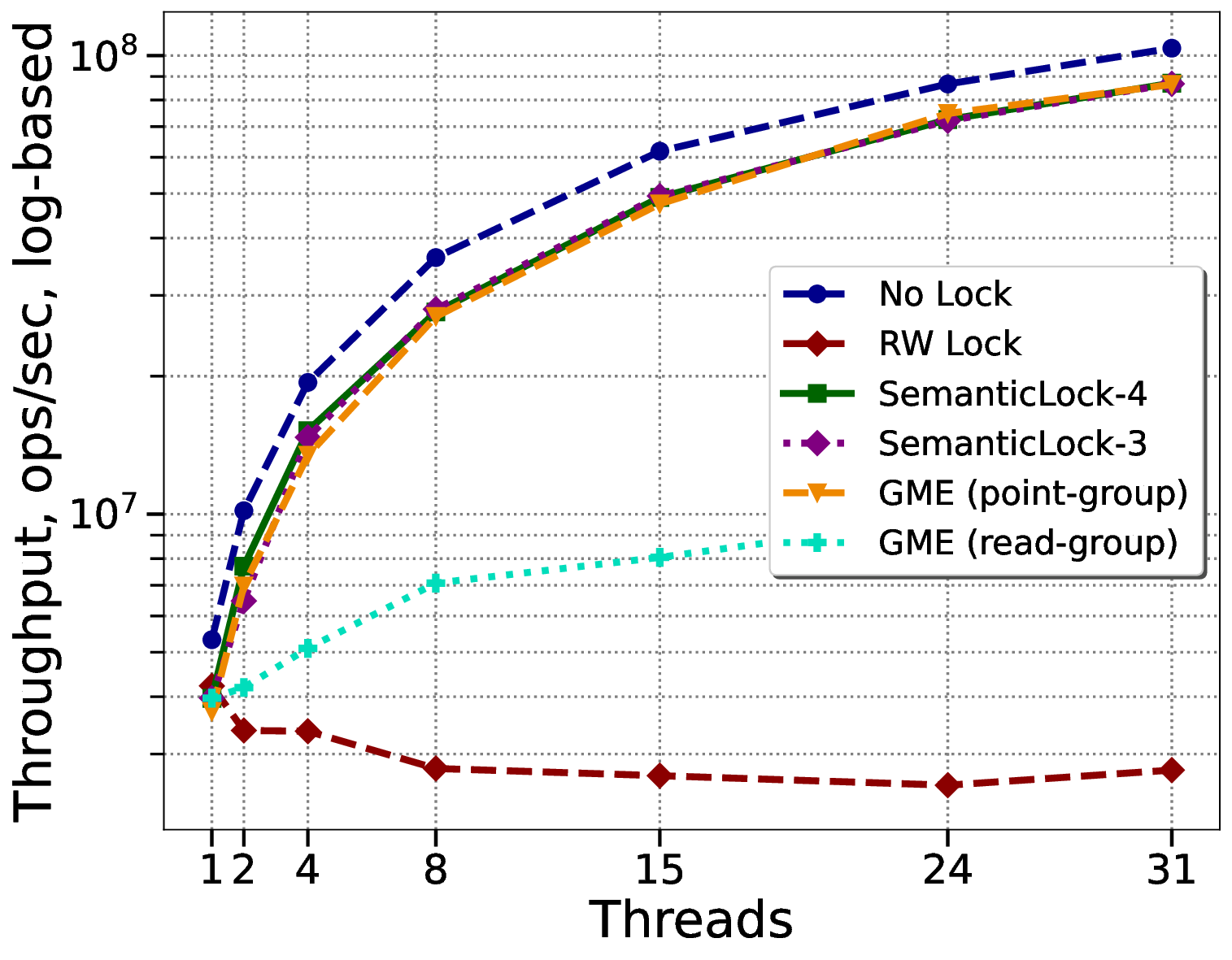}
			\small (a) $50\%$ read, $50\%$ modify
		\end{minipage}
        \begin{minipage}[t]{0.32\linewidth}
			\centering
			\includegraphics[width=1.0\linewidth]
            {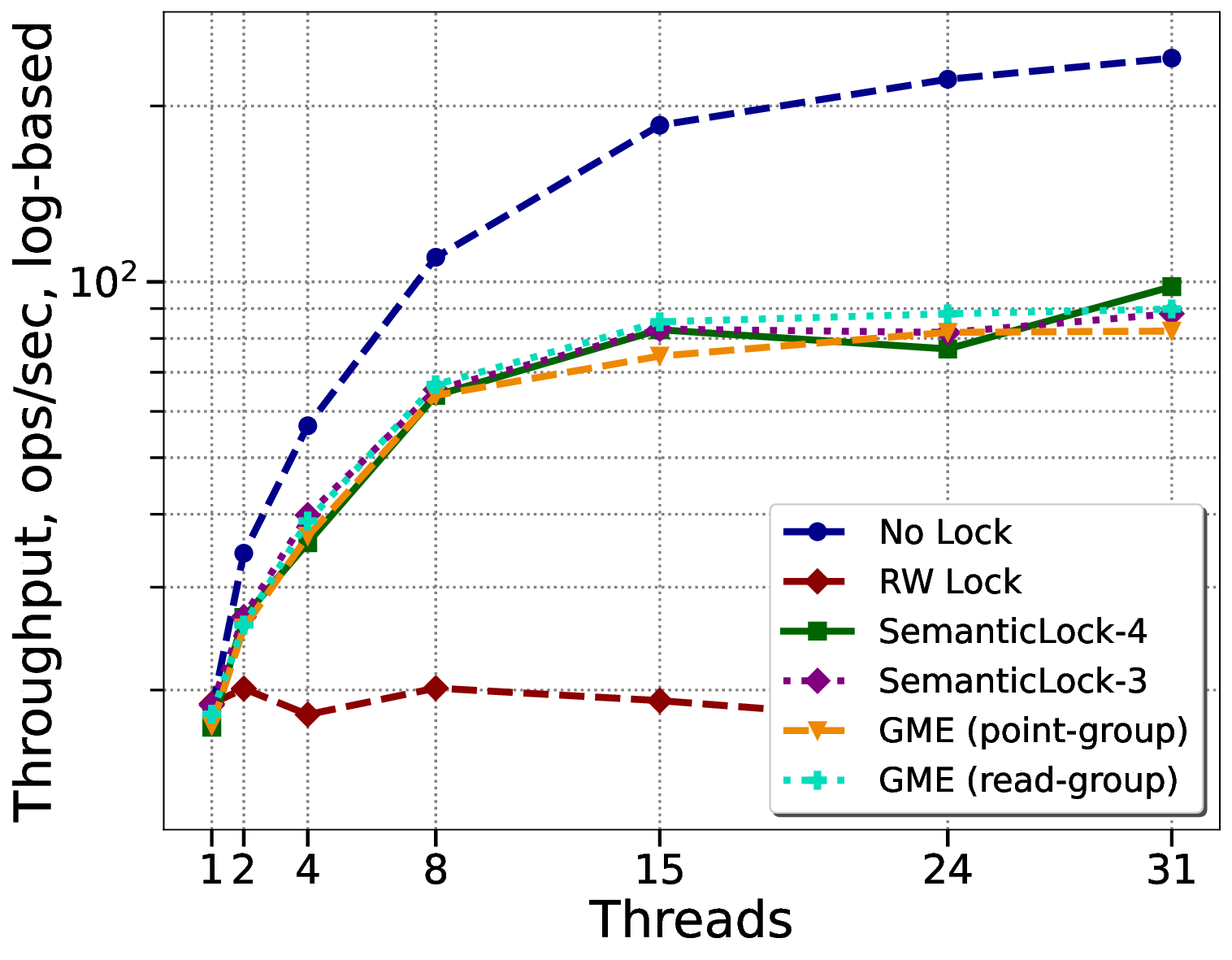}
			\small (b) $50\%$ read, $50\%$ keysSnapshot
		\end{minipage}
        \begin{minipage}[t]{0.32\linewidth}
			\centering
			\includegraphics[width=1.0\linewidth]
            {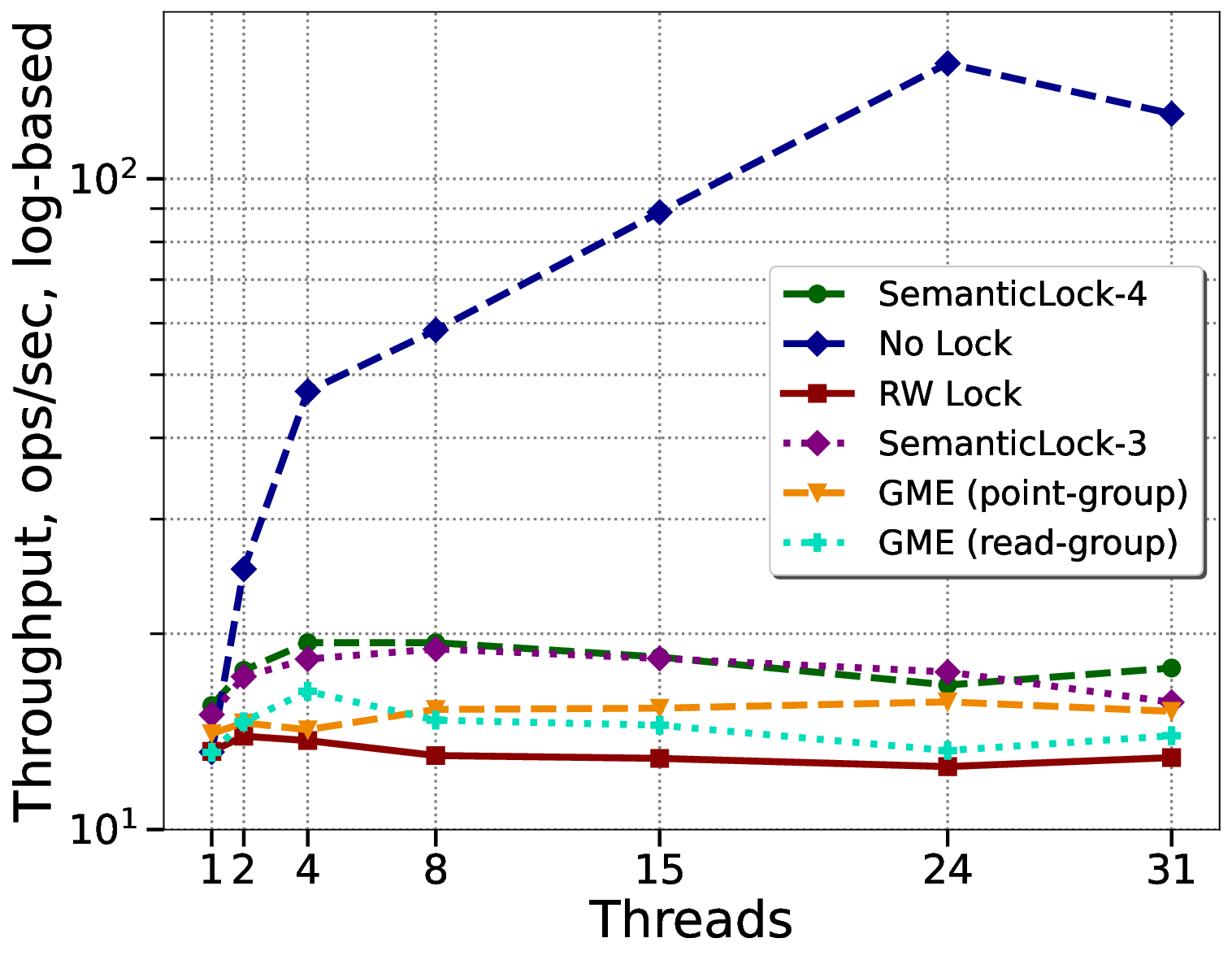}
			\small (c) $25\%$ read, $25\%$ modify, \\$25\%$ keysS-t, $25\%$ mapValues
		\end{minipage}
	\end{minipage}
\begin{minipage}[t]{1.0\linewidth}
		\centering
		\begin{minipage}[t]{0.32\linewidth}
			\centering
			\includegraphics[width=1.0\linewidth]
            {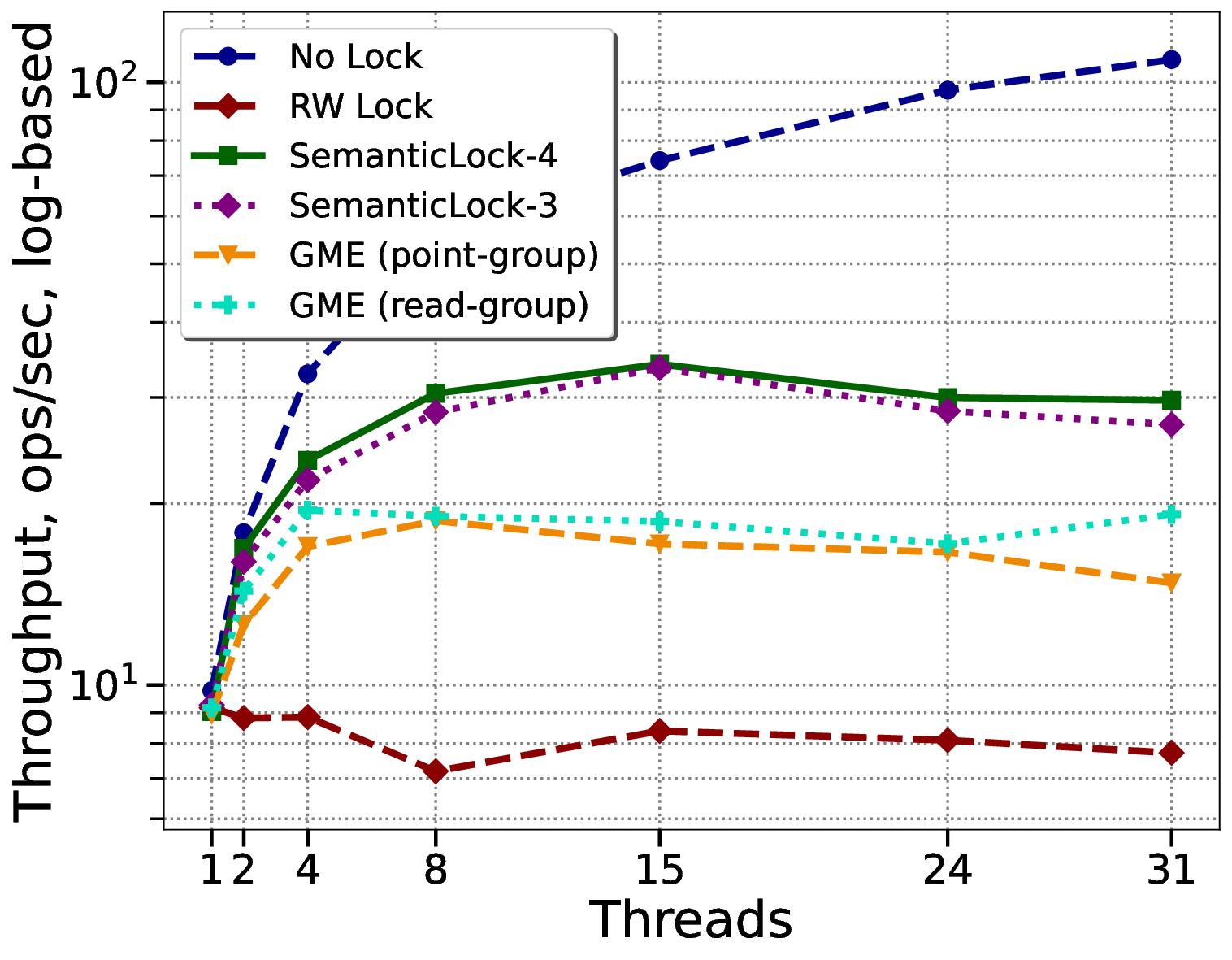}
			\small (d) $5\%$ read, $5\%$ modify, \\$80\%$ keysS-t, $10\%$ mapValues
		\end{minipage}
        \begin{minipage}[t]{0.32\linewidth}
			\centering
			\includegraphics[width=1.0\linewidth]
            {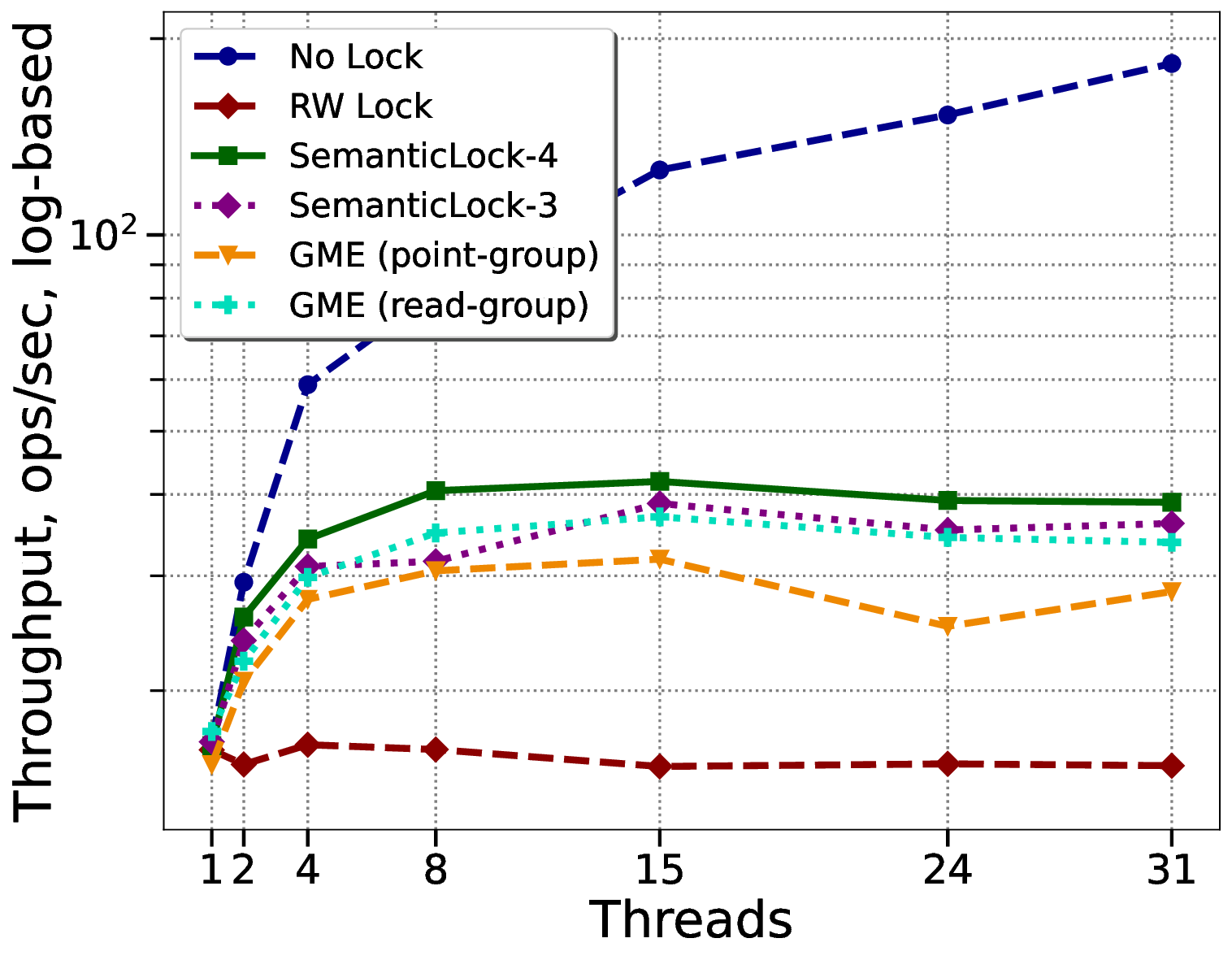}
			\small (e) $10\%$ read, $40\%$ modify, \\$45\%$ keysS-t, $5\%$ mapValues
		\end{minipage}
        \begin{minipage}[t]{0.32\linewidth}
			\centering
			\includegraphics[width=1.0\linewidth]
            {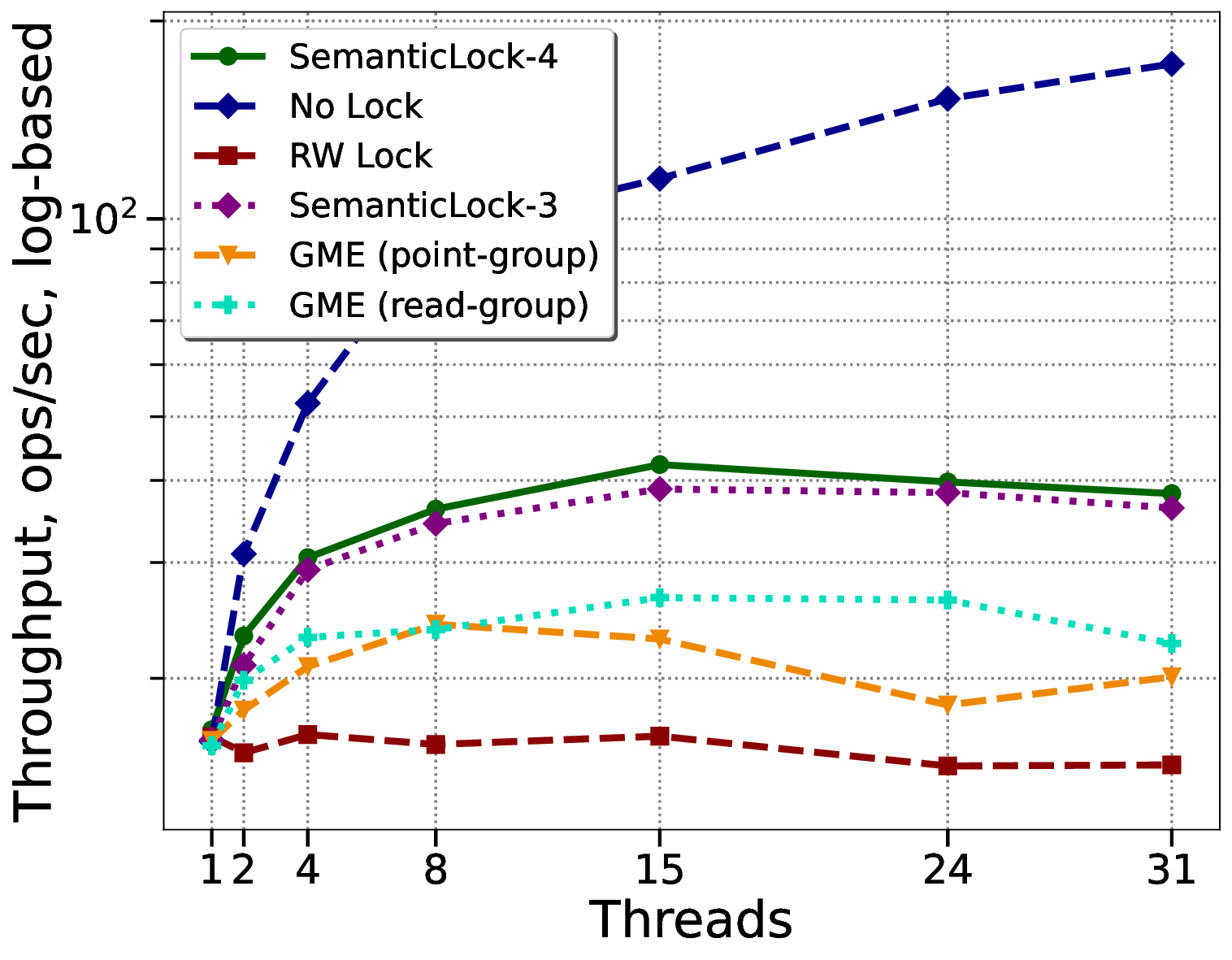}
			\small (f) $40\%$ read, $10\%$ modify, \\$40\%$ keysS-t, $10\%$ mapValues
		\end{minipage}
	\end{minipage}

	\caption{Throughput of extended ConcurrentHashMap under varying operation distributions.}
	\label{fig:map}
\end{figure}

Figure~\ref{fig:map} reports the results for the extended \texttt{ConcurrentHashMap}. The trends are consistent with the array benchmark: for workloads consisting only of standard point operations or read-only operations, \texttt{SemanticLock} incurs additional synchronization overhead relative to the \texttt{No Lock} upper bound, but preserves high scalability.

Both \texttt{SemanticLock} variants scale until hyper-threading effects become visible at higher thread counts. The fine-grained graph used by \texttt{SemanticLock-4} provides a modest additional benefit by allowing \textsf{get} to overlap with \textsf{keysSnapshot} while still excluding point updates.

\bibliography{bib}

@inproceedings{Aksenov2019,
  author = {Aksenov, Vitaly and Kuznetsov, Petr and Shalyto, Anatoly},
  title = {{Parallel Combining: Benefits of Explicit Synchronization}},
  booktitle = {22nd International Conference on Principles of Distributed Systems (OPODIS 2018)},
  pages = {11:1--11:16},
  series = {Leibniz International Proceedings in Informatics (LIPIcs)},
  isbn = {978-3-95977-098-9},
  issn = {1868-8969},
  year = {2019},
  volume = {125},
  publisher = {Schloss Dagstuhl -- Leibniz-Zentrum f{\"u}r Informatik},
  address = {Dagstuhl, Germany},
  url = {https://drops.dagstuhl.de/entities/document/10.4230/LIPIcs.OPODIS.2018.11},
  urn = {urn:nbn:de:0030-drops-100713},
  doi = {10.4230/LIPIcs.OPODIS.2018.11},
  annote = {Keywords: concurrent data structure, parallel batched data structure, combining},
  timestamp = {Tue, 15 Feb 2022 00:00:00 +0100},
  biburl = {https://dblp.org/rec/conf/opodis/AksenovKS18.bib},
  bibsource = {dblp computer science bibliography, https://dblp.org},
  _bib2doi_selected = {dblp:/rec/conf/opodis/AksenovKS18.bib},
  _bib2doi_confirmed = {true},
}

@inproceedings{Herlihy_Boosting,
  author = {Herlihy, Maurice and Koskinen, Eric},
  title = {Transactional boosting: a methodology for highly-concurrent transactional objects},
  year = {2008},
  isbn = {9781595937957},
  publisher = {Association for Computing Machinery},
  address = {New York, NY, USA},
  url = {https://doi.org/10.1145/1345206.1345237},
  doi = {10.1145/1345206.1345237},
  abstract = {We describe a methodology for transforming a large class of highly-concurrent linearizable objects into highly-concurrent transactional objects. As long as the linearizable implementation satisfies certain regularity properties (informally, that every method has an inverse), we define a simple wrapper for the linearizable implementation that guarantees that concurrent transactions without inherent conflicts can synchronize at the same granularity as the original linearizable implementation.},
  booktitle = {Proceedings of the 13th ACM SIGPLAN Symposium on Principles and Practice of Parallel Programming},
  pages = {207–216},
  numpages = {10},
  keywords = {transactional memory, transactional boosting, non-blocking algorithms, commutativity, abstract locks},
  location = {Salt Lake City, UT, USA},
  series = {PPoPP '08},
  timestamp = {Wed, 25 Feb 2026 00:00:00 +0100},
  biburl = {https://dblp.org/rec/conf/ppopp/HerlihyK08.bib},
  bibsource = {dblp computer science bibliography, https://dblp.org},
  _bib2doi_selected = {dblp:/rec/conf/ppopp/HerlihyK08.bib},
  _bib2doi_confirmed = {true},
  _bib2doi_finished = {true},
}

@inproceedings{BrownCollaborative,
  author = {Brown, Trevor and Prokopec, Aleksandar and Alistarh, Dan},
  title = {Non-blocking interpolation search trees with doubly-logarithmic running time},
  year = {2020},
  isbn = {9781450368186},
  publisher = {Association for Computing Machinery},
  address = {New York, NY, USA},
  url = {https://doi.org/10.1145/3332466.3374542},
  doi = {10.1145/3332466.3374542},
  abstract = {Balanced search trees typically use key comparisons to guide their operations, and achieve logarithmic running time. By relying on numerical properties of the keys, interpolation search achieves lower search complexity and better performance. Although interpolation-based data structures were investigated in the past, their non-blocking concurrent variants have received very little attention so far.In this paper, we propose the first non-blocking implementation of the classic interpolation search tree (IST) data structure. For arbitrary key distributions, the data structure ensures worst-case O(log n + p) amortized time for search, insertion and deletion traversals. When the input key distributions are smooth, lookups run in expected O(log log n + p) time, and insertion and deletion run in expected amortized O(log log n + p) time, where p is a bound on the number of threads. To improve the scalability of concurrent insertion and deletion, we propose a novel parallel rebuilding technique, which should be of independent interest.We evaluate whether the theoretical improvements translate to practice by implementing the concurrent interpolation search tree, and benchmarking it on uniform and nonuniform key distributions, for dataset sizes in the millions to billions of keys. Relative to the state-of-the-art concurrent data structures, the concurrent interpolation search tree achieves performance improvements of up to 15\% under high update rates, and of up to 50\% under moderate update rates. Further, ISTs exhibit up to 2X less cache-misses, and consume 1.2 -- 2.6X less memory compared to the next best alternative on typical dataset sizes. We find that the results are surprisingly robust to distributional skew, which suggests that our data structure can be a promising alternative to classic concurrent search structures.},
  booktitle = {Proceedings of the 25th ACM SIGPLAN Symposium on Principles and Practice of Parallel Programming},
  pages = {276–291},
  numpages = {16},
  keywords = {search trees, non-blocking algorithms, interpolation, concurrent data structures},
  location = {San Diego, California},
  series = {PPoPP '20},
  timestamp = {Sun, 12 Jun 2022 01:00:00 +0200},
  biburl = {https://dblp.org/rec/conf/ppopp/BrownPA20.bib},
  bibsource = {dblp computer science bibliography, https://dblp.org},
  _bib2doi_selected = {dblp:/rec/conf/ppopp/BrownPA20.bib},
  _bib2doi_confirmed = {true},
  _bib2doi_finished = {true},
}

@inproceedings{Verlib,
  author = {Blelloch, Guy E. and Wei, Yuanhao},
  title = {VERLIB: Concurrent Versioned Pointers},
  year = {2024},
  isbn = {9798400704352},
  publisher = {Association for Computing Machinery},
  address = {New York, NY, USA},
  url = {https://doi.org/10.1145/3627535.3638501},
  doi = {10.1145/3627535.3638501},
  booktitle = {Proceedings of the 29th ACM SIGPLAN Annual Symposium on Principles and Practice of Parallel Programming},
  pages = {200–214},
  numpages = {15},
  keywords = {multiversioning, concurrent data structures, snapshots, lock-based, lock-free},
  location = {Edinburgh, United Kingdom},
  series = {PPoPP '24},
  timestamp = {Mon, 01 Apr 2024 01:00:00 +0200},
  biburl = {https://dblp.org/rec/conf/ppopp/BlellochW24.bib},
  bibsource = {dblp computer science bibliography, https://dblp.org},
  _bib2doi_selected = {dblp:/rec/conf/ppopp/BlellochW24.bib},
  _bib2doi_confirmed = {true},
  _bib2doi_finished = {true},
}

@inproceedings{Lomet,
  author = {Lomet, David B.},
  title = {Key Range Locking Strategies for Improved Concurrency},
  year = {1993},
  isbn = {155860152X},
  publisher = {Morgan Kaufmann Publishers Inc.},
  address = {San Francisco, CA, USA},
  booktitle = {Proceedings of the 19th International Conference on Very Large Data Bases},
  pages = {655–664},
  numpages = {10},
  series = {VLDB '93},
  timestamp = {Tue, 07 Nov 2017 16:24:37 +0100},
  biburl = {https://dblp.org/rec/conf/vldb/Lomet93.bib},
  bibsource = {dblp computer science bibliography, https://dblp.org},
  url = {http://www.vldb.org/conf/1993/P655.PDF},
  _bib2doi_selected = {dblp:/rec/conf/vldb/Lomet93.bib},
  _bib2doi_confirmed = {true},
  _bib2doi_finished = {true},
}

@article{Badri,
  author = {B. R. Badrinath and Krithi Ramamritham},
  title = {Semantics-Based Concurrency Control: Beyond Commutativity},
  journal = {{ACM} Trans. Database Syst.},
  volume = {17},
  number = {1},
  pages = {163--199},
  year = {1992},
  url = {https://doi.org/10.1145/128765.128771},
  doi = {10.1145/128765.128771},
  timestamp = {Tue, 06 Nov 2018 00:00:00 +0100},
  biburl = {https://dblp.org/rec/journals/tods/BadrinathR92.bib},
  bibsource = {dblp computer science bibliography, https://dblp.org},
  _bib2doi_selected = {dblp:/rec/journals/tods/BadrinathR92.bib},
  _bib2doi_confirmed = {true},
}

@article{Linearizability,
  author = {Herlihy, Maurice P. and Wing, Jeannette M.},
  title = {Linearizability: a correctness condition for concurrent objects},
  year = {1990},
  issue_date = {July 1990},
  publisher = {Association for Computing Machinery},
  address = {New York, NY, USA},
  volume = {12},
  number = {3},
  issn = {0164-0925},
  url = {https://doi.org/10.1145/78969.78972},
  doi = {10.1145/78969.78972},
  journal = {ACM Trans. Program. Lang. Syst.},
  month = {jul},
  pages = {463–492},
  numpages = {30},
  timestamp = {Wed, 14 Nov 2018 00:00:00 +0100},
  biburl = {https://dblp.org/rec/journals/toplas/HerlihyW90.bib},
  bibsource = {dblp computer science bibliography, https://dblp.org},
  _bib2doi_selected = {dblp:/rec/journals/toplas/HerlihyW90.bib},
  _bib2doi_confirmed = {true},
  _bib2doi_finished = {true},
}

@inproceedings{Lincheck,
  author = {Potapov, Aleksandr and Zuev, Maksim and Moiseenko, Evgenii and Koval, Nikita},
  title = {Testing Concurrent Algorithms on JVM with Lincheck and IntelliJ IDEA},
  year = {2024},
  isbn = {9798400706127},
  publisher = {Association for Computing Machinery},
  address = {New York, NY, USA},
  url = {https://doi.org/10.1145/3650212.3685301},
  doi = {10.1145/3650212.3685301},
  booktitle = {Proceedings of the 33rd ACM SIGSOFT International Symposium on Software Testing and Analysis},
  pages = {1821–1825},
  numpages = {5},
  keywords = {IDE, concurrency, model checking, record-and-replay debugging},
  location = {Vienna, Austria},
  series = {ISSTA 2024},
  timestamp = {Sun, 19 Jan 2025 00:00:00 +0100},
  biburl = {https://dblp.org/rec/conf/issta/PotapovZMK24.bib},
  bibsource = {dblp computer science bibliography, https://dblp.org},
  _bib2doi_selected = {dblp:/rec/conf/issta/PotapovZMK24.bib},
  _bib2doi_confirmed = {true},
  _bib2doi_finished = {true},
}

@inproceedings{Synchrobench,
  author = {Gramoli, Vincent},
  title = {More than you ever wanted to know about synchronization: synchrobench, measuring the impact of the synchronization on concurrent algorithms},
  year = {2015},
  isbn = {9781450332057},
  publisher = {Association for Computing Machinery},
  address = {New York, NY, USA},
  url = {https://doi.org/10.1145/2688500.2688501},
  doi = {10.1145/2688500.2688501},
  booktitle = {Proceedings of the 20th ACM SIGPLAN Symposium on Principles and Practice of Parallel Programming},
  pages = {1–10},
  numpages = {10},
  keywords = {Benchmark, data structure, lock-freedom, reusability},
  location = {San Francisco, CA, USA},
  series = {PPoPP 2015},
  timestamp = {Sun, 12 Jun 2022 01:00:00 +0200},
  biburl = {https://dblp.org/rec/conf/ppopp/Gramoli15.bib},
  bibsource = {dblp computer science bibliography, https://dblp.org},
  _bib2doi_selected = {dblp:/rec/conf/ppopp/Gramoli15.bib},
  _bib2doi_confirmed = {true},
  _bib2doi_finished = {true},
}

@inproceedings{GME,
  author = {Joung, Yuh-Jzer},
  title = {Asynchronous group mutual exclusion (extended abstract)},
  year = {1998},
  isbn = {0897919777},
  publisher = {Association for Computing Machinery},
  address = {New York, NY, USA},
  url = {https://doi.org/10.1145/277697.277706},
  doi = {10.1145/277697.277706},
  booktitle = {Proceedings of the Seventeenth Annual ACM Symposium on Principles of Distributed Computing},
  pages = {51–60},
  numpages = {10},
  location = {Puerto Vallarta, Mexico},
  series = {PODC '98},
  timestamp = {Tue, 06 Nov 2018 00:00:00 +0100},
  biburl = {https://dblp.org/rec/conf/podc/Joung98.bib},
  bibsource = {dblp computer science bibliography, https://dblp.org},
  _bib2doi_selected = {dblp:/rec/conf/podc/Joung98.bib},
  _bib2doi_confirmed = {true},
  _bib2doi_finished = {true},
}

@article{RW,
  author = {Courtois, P. J. and Heymans, F. and Parnas, D. L.},
  title = {Concurrent control with “readers” and “writers”},
  year = {1971},
  issue_date = {Oct. 1971},
  publisher = {Association for Computing Machinery},
  address = {New York, NY, USA},
  volume = {14},
  number = {10},
  issn = {0001-0782},
  url = {https://doi.org/10.1145/362759.362813},
  doi = {10.1145/362759.362813},
  journal = {Commun. ACM},
  month = {oct},
  pages = {667–668},
  numpages = {2},
  keywords = {shared access to resources, mutual exclusion, critical section},
  timestamp = {Tue, 06 Nov 2018 00:00:00 +0100},
  biburl = {https://dblp.org/rec/journals/cacm/CouroisHP71.bib},
  bibsource = {dblp computer science bibliography, https://dblp.org},
  _bib2doi_selected = {dblp:/rec/journals/cacm/CouroisHP71.bib},
  _bib2doi_confirmed = {true},
  _bib2doi_finished = {true},
}

@inproceedings{hendler2010flat,
  author = {Hendler, Danny and Incze, Itai and Shavit, Nir and Tzafrir, Moran},
  year = {2010},
  month = {06},
  pages = {355-364},
  title = {Flat combining and the synchronization-parallelism tradeoff},
  booktitle = {Annual ACM Symposium on Parallelism in Algorithms and Architectures},
  doi = {10.1145/1810479.1810540},
  timestamp = {Tue, 06 Nov 2018 00:00:00 +0100},
  biburl = {https://dblp.org/rec/conf/spaa/HendlerIST10.bib},
  bibsource = {dblp computer science bibliography, https://dblp.org},
  _bib2doi_selected = {dblp:/rec/conf/spaa/HendlerIST10.bib},
  _bib2doi_confirmed = {true},
}

@inproceedings{shavit1995software,
  title = {Software transactional memory},
  author = {Shavit, Nir and Touitou, Dan},
  booktitle = {Proceedings of the fourteenth annual ACM symposium on Principles of distributed computing},
  pages = {204--213},
  year = {1995},
  timestamp = {Tue, 06 Nov 2018 00:00:00 +0100},
  biburl = {https://dblp.org/rec/conf/podc/ShavitT95.bib},
  bibsource = {dblp computer science bibliography, https://dblp.org},
  doi = {10.1145/224964.224987},
  _bib2doi_selected = {dblp:/rec/conf/podc/ShavitT95.bib},
  _bib2doi_confirmed = {true},
}

@inproceedings{rodriguez2025skip,
  title = {Skip Hash: A Fast Ordered Map Via Software Transactional Memory},
  author = {Rodriguez, Matthew and Aksenov, Vitaly and Spear, Michael},
  booktitle = {2025 IEEE 45th International Conference on Distributed Computing Systems (ICDCS)},
  pages = {956--966},
  year = {2025},
  organization = {IEEE},
  doi = {10.1109/ICDCS63083.2025.00097},
  timestamp = {Thu, 25 Dec 2025 00:00:00 +0100},
  biburl = {https://dblp.org/rec/conf/icdcs/RodriguezAS25.bib},
  bibsource = {dblp computer science bibliography, https://dblp.org},
  _bib2doi_selected = {dblp:/rec/conf/icdcs/RodriguezAS25.bib},
  _bib2doi_confirmed = {true},
}

@inproceedings{Luo2013LGME,
  author = {Luo, Aoxue and Wu, Weigang and Cao, Jiannong and Raynal, Michel},
  year = {2013},
  pages = {300-309},
  title = {A Generalized Mutual Exclusion Problem and Its Algorithm},
  booktitle = {Proceedings of the International Conference on Parallel Processing},
  doi = {10.1109/ICPP.2013.39},
  timestamp = {Fri, 24 Mar 2023 00:00:00 +0100},
  biburl = {https://dblp.org/rec/conf/icpp/AoxueluoWCR13.bib},
  bibsource = {dblp computer science bibliography, https://dblp.org},
  _bib2doi_selected = {dblp:/rec/conf/icpp/AoxueluoWCR13.bib},
  _bib2doi_confirmed = {true},
}

@inproceedings{Dice2013,
  author = {Dave Dice and Yossi Lev and Mark Moir},
  title = {Scalable statistics counters},
  booktitle = {25th {ACM} Symposium on Parallelism in Algorithms and Architectures, {SPAA} '13},
  pages = {43--52},
  publisher = {{ACM}},
  year = {2013},
  doi = {10.1145/2486159.2486182},
  timestamp = {Tue, 06 Nov 2018 00:00:00 +0100},
  biburl = {https://dblp.org/rec/conf/spaa/DiceLM13.bib},
  bibsource = {dblp computer science bibliography, https://dblp.org},
  _bib2doi_selected = {dblp:/rec/conf/spaa/DiceLM13.bib},
  _bib2doi_confirmed = {true},
}

@inproceedings{MML,
  author = {Kulkarni, Milind and Nguyen, Donald and Prountzos, Dimitrios and Sui, Xin and Pingali, Keshav},
  year = {2011},
  month = {06},
  pages = {542-555},
  title = {Exploiting the Commutativity Lattice},
  volume = {46},
  booktitle = {ACM SIGPLAN Notices},
  doi = {10.1145/1993498.1993562},
  timestamp = {Thu, 24 Jun 2021 01:00:00 +0200},
  biburl = {https://dblp.org/rec/conf/pldi/KulkarniNPSP11.bib},
  bibsource = {dblp computer science bibliography, https://dblp.org},
  _bib2doi_selected = {dblp:/rec/conf/pldi/KulkarniNPSP11.bib},
  _bib2doi_confirmed = {true},
}

\appendix
\section{SemanticLock Correctness and Fairness}
\label{app:semanticlock-correctness-fairness}

\subsection{Correctness Proof}
\label{app:semanticlock-correctness}
We assume that the given conflict graph is \emph{sound}: every pair of operation types whose overlapping executions may violate correctness is connected by an edge, and every operation type that is not safe to execute concurrently with itself has a self-loop.

We first show that SemanticLock never admits two conflicting operations simultaneously. Consider two operation instances $a$ and $b$ of types $u$ and $v$ such that $(u,v)\in E$; the case $u=v$ corresponds to a self-loop. Suppose, for contradiction, that both instances successfully acquire SemanticLock and their protected executions overlap. Let $reserve(x)$ denote the successful \texttt{LongAdder} increment or \texttt{AtomicInteger} compare-and-set by which operation instance $x$ reserves its type, let $validate(x)$ denote the validation step of $x$, and let $release(x)$ denote the decrement performed when $x$ leaves the protected region.

Without loss of generality, $reserve(a)$ precedes $reserve(b)$ in the linearization order of atomic counter operations. Since the protected executions of $a$ and $b$ overlap, $a$ does not execute $release(a)$ before $b$ completes its validation; otherwise, $a$ would have already left the protected region before $b$ entered it. Therefore, during $validate(b)$, the counter $cnt[u]$ is positive. Because $u\in N(v)$ when $u\ne v$, and because a self-loop is handled by the compare-and-set reservation when $u=v$, operation $b$ cannot successfully validate while $a$ remains reserved. It must either fail the compare-and-set in the self-loop case or observe a positive conflicting counter and roll back. This contradicts the assumption that both operations acquire successfully and overlap. So, for every edge $(u,v)\in E$, counters of adjacent conflicting types are never positive simultaneously after successful acquisition.

It remains to connect this admission property to the correctness of the wrapped data structure. By the soundness assumption on the conflict graph, all pairs of operations whose overlapping executions may violate correctness are represented by edges or self-loops. SemanticLock excludes exactly such overlaps, while allowing only pairs of operations that are safe to execute concurrently according to the graph. Therefore, if each operation implementation is linearizable when executed alone or in any graph-permitted overlap, wrapping all operations with SemanticLock preserves linearizability of the resulting data structure.

\subsection{Optional Fairness Guarantee}
\label{app:semanticlock-fairness}
The optimistic verification protocol guarantees correctness, but not progress. In particular, an operation of type $u$ may starve if a continuous stream of conflicting operations repeatedly acquires SemanticLock before $u$ manages to reserve and validate its counter. We therefore provide an optional fairness mode.

In fairness mode, SemanticLock maintains a concurrent list of pending requests. A request record contains the thread identifier and the requested operation type. Before starting verification, a thread appends its record to the tail of the list. The verification phase is modified as follows: during \textit{precheck}, a request of type $u$ may proceed only if two conditions hold. First, all currently executing conflicting operations are absent, as in the basic precheck. Second, there is no earlier request in the list whose type conflicts with $u$. If either condition fails, the thread keeps its record in the list and retries later. Once the request successfully passes the order check, it saves that fact in the record and does not check it again because no requests can appear earlier in the list. Once the request successfully passes \textit{validation} and enters the protected region, it removes its record from the list; the operation itself is still protected by the counter $cnt[u]$ until release.

This rule allows non-conflicting requests to bypass one another, but it prevents a later conflicting request from overtaking an earlier one. Consider a request $r$ of type $u$ after it has been enqueued. Any conflicting request that is enqueued after $r$ cannot pass \textit{precheck} before $r$ is removed from the list, because it observes $r$ as an earlier conflicting request. Thus, only conflicting requests that were already before $r$ in the list, together with conflicting operations already executing at the time of enqueue, can delay $r$. This set is finite under a finite number of threads. After those operations complete or enter and leave the protected region, no later conflicting request can bypass $r$, so eventually $r$ observes both no earlier conflicting request and no executing conflicting operation, passes verification, and enters the protected region. Hence, the fairness mode provides starvation-freedom for conflicting operations; moreover, the number of conflicting bypasses is bounded by the number of requests already ahead of $r$ when it is enqueued.

The fairness mode is optional and can be enabled by a parameter when creating a SemanticLock instance.

\section{Extra Experiments with Varying Fractions of Write and Read Operations}
\label{app:extra-exp-writes}

\begin{figure}[!ht]
	\begin{minipage}[t]{1.0\linewidth}
		\centering
		\begin{minipage}[t]{0.32\linewidth}
			\centering
			\includegraphics[width=1.0\linewidth]
            {pic/2map-50-0-50-gme.eps}
			\small (a) $50\%$ read, $0\%$ modify, \\$50\%$ keysS-t, $0\%$ mapValues
		\end{minipage}
        \begin{minipage}[t]{0.32\linewidth}
			\centering
			\includegraphics[width=1.0\linewidth]
            {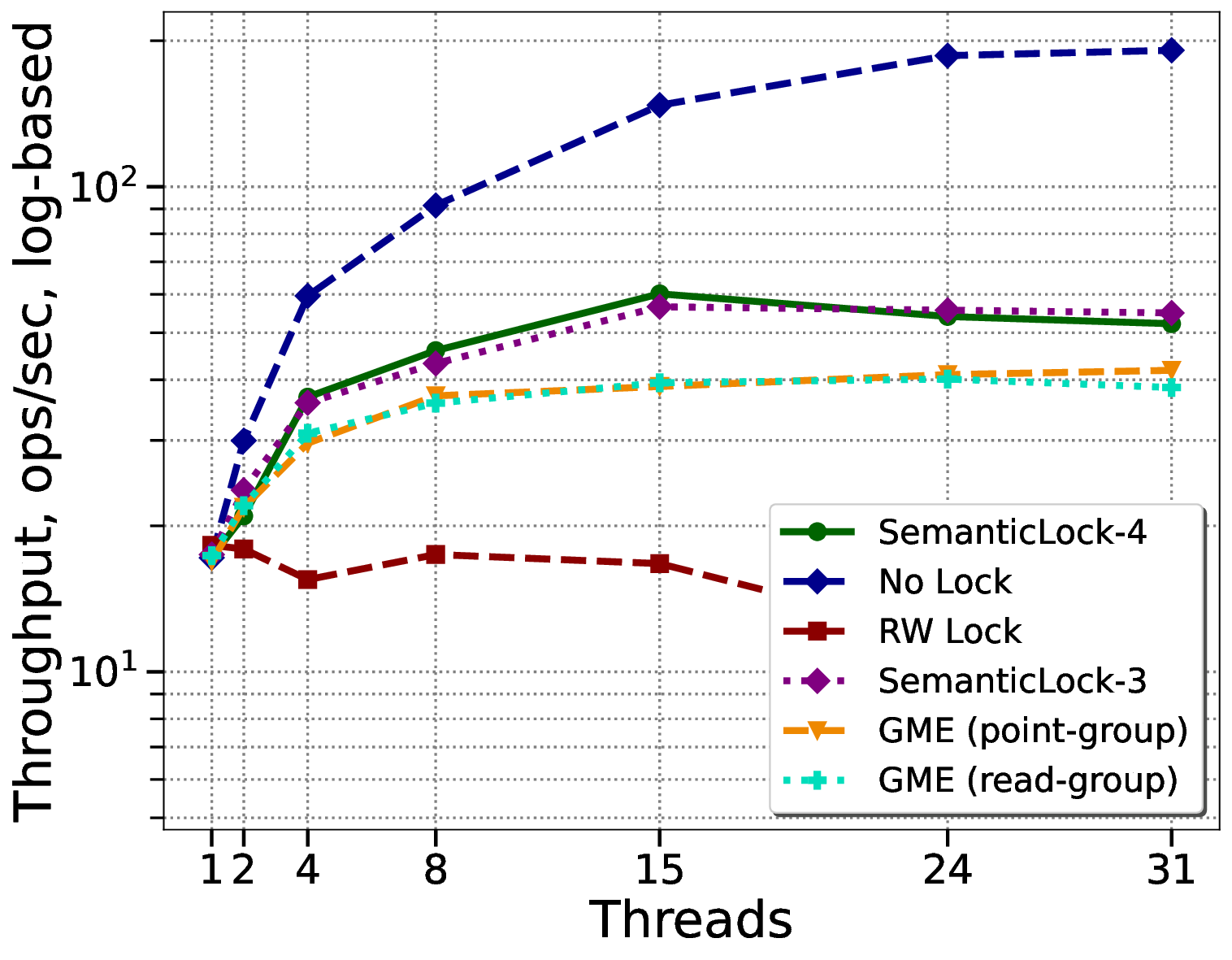}
			\small (b) $45\%$ read, $5\%$ modify, \\$45\%$ keysS-t, $5\%$ mapValues
		\end{minipage}
        \begin{minipage}[t]{0.32\linewidth}
			\centering
			\includegraphics[width=1.0\linewidth]
            {pic/2map-40-10-40-gme.eps}
			\small (c) $40\%$ read, $10\%$ modify, \\$40\%$ keysS-t, $10\%$ mapValues
		\end{minipage}
	\end{minipage}
\begin{minipage}[t]{1.0\linewidth}
		\centering
		\begin{minipage}[t]{0.32\linewidth}
			\centering
			\includegraphics[width=1.0\linewidth]
            {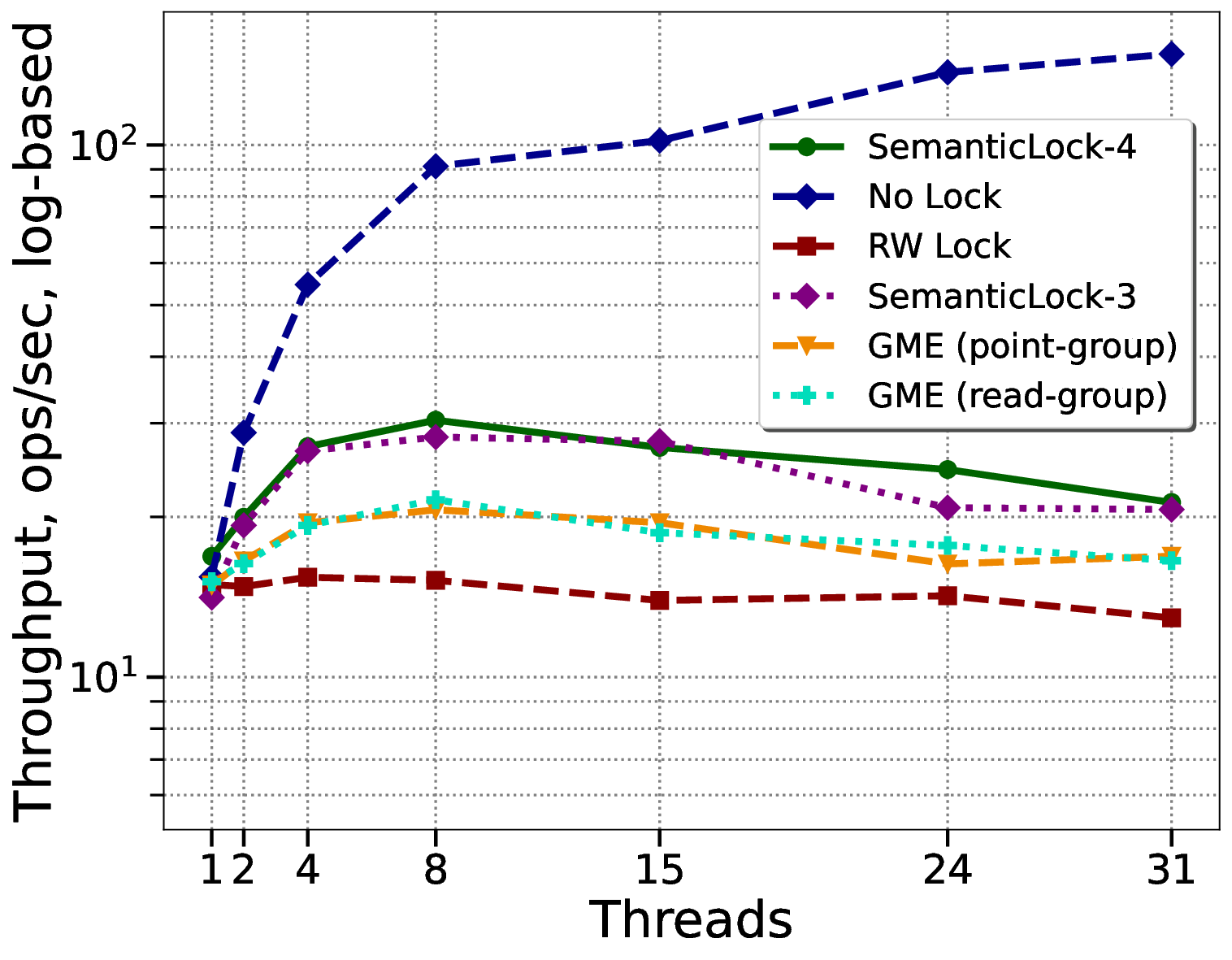}
			\small (d) $35\%$ read, $15\%$ modify, \\$35\%$ keysS-t, $15\%$ mapValues
		\end{minipage}
        \begin{minipage}[t]{0.32\linewidth}
			\centering
			\includegraphics[width=1.0\linewidth]
            {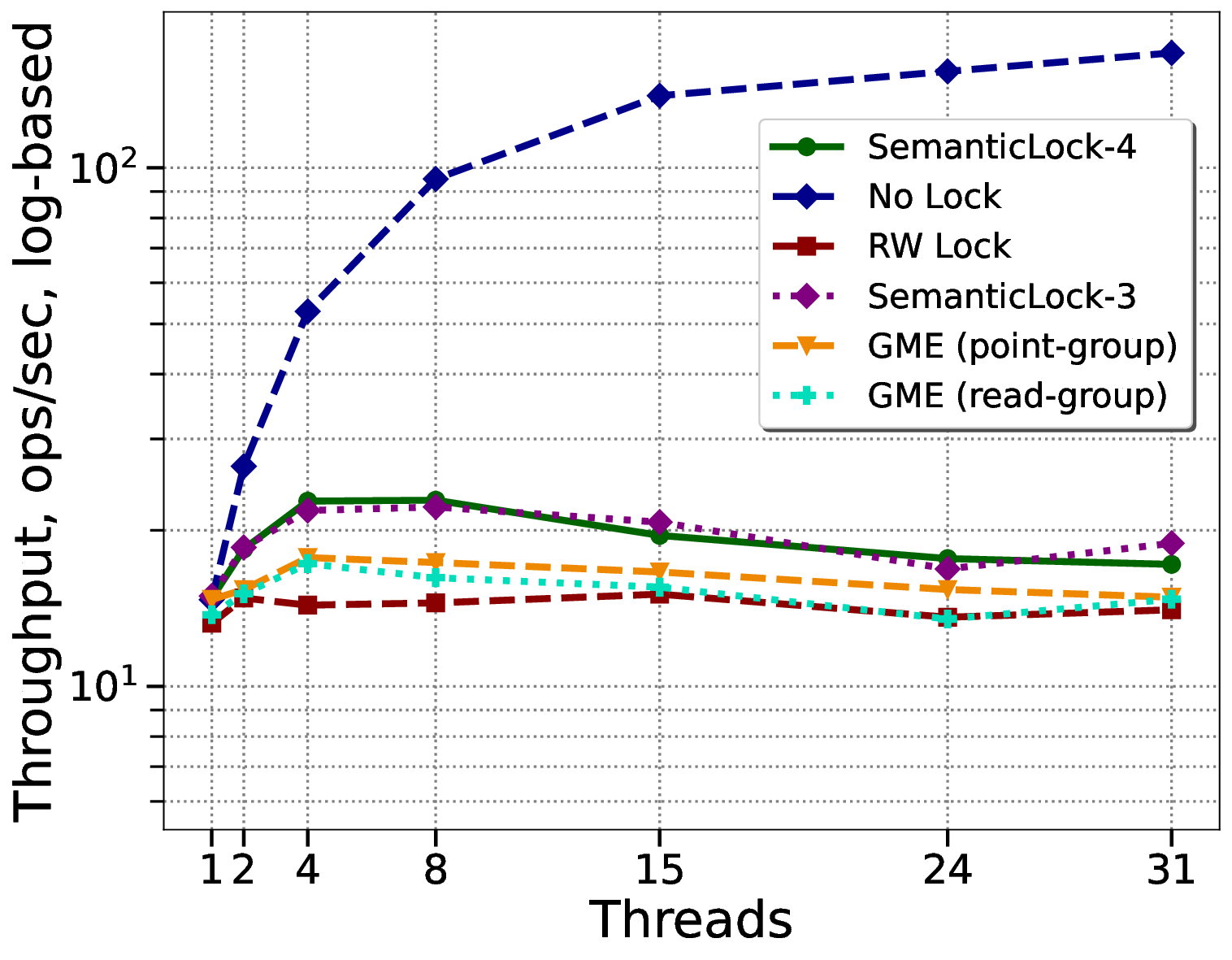}
			\small (e) $30\%$ read, $20\%$ modify, \\$30\%$ keysS-t, $20\%$ mapValues
		\end{minipage}
        \begin{minipage}[t]{0.32\linewidth}
			\centering
			\includegraphics[width=1.0\linewidth]
            {pic/2map-25-25-25-gme.eps}
			\small (f) $25\%$ read, $25\%$ modify, \\$25\%$ keysS-t, $25\%$ mapValues
		\end{minipage}
	\end{minipage}

	\caption{Throughput of extended ConcurrentHashMap under varying fractions of write operations}
	\label{fig:map-writes}
\end{figure}

We additionally run experiments on the changing proportion of write operations (Figure~\ref{fig:map-writes}). 

As expected, the advantage of SemanticLock decreases as writes become more frequent. Once writes and reads occur at approximately equal rates, SemanticLock no longer scales with increasing thread count and provides no meaningful throughput improvement over the baselines. The intended use case is therefore a workload in which the conflict graph contains substantial compatibility not expressible by an RW lock or GME. SemanticLock is not expected to help when most operations conflict.

\section{Atomic Increments in Java}
\label{app:atomic-problem}
Initially, we started with an implementation with only \texttt{AtomicIntegers}. It worked in the same way as our current implementation but had very low scalability in scenarios with only point operations. The reason was a appearance of large number of conflicts during \texttt{incrementAndGet} of \texttt{AtomicInteger} and its low scalability. To make sure that this is a problem, we removed all other code from the \texttt{lock} method and simply incremented the counter there. The scalability remained low.

To mitigate the severe overhead associated with heavily contended atomic operations, modern concurrent systems frequently employ striped counters~\cite{Dice2013}. In Java, this idea is used in \texttt{LongAdder}. Rather than forcing concurrent threads to contend on a single memory location via single-point compare-and-set instructions, \texttt{LongAdder} dynamically distributes updates across an array of independent memory structures, internally referred to as cells. Under high contention, threads utilize a thread-specific hash code to route their increment operations to distinct cells, effectively transforming a global contention bottleneck into multiple parallel, uncontended updates.

After changing \texttt{AtomicInteger} to \texttt{LongAdder} for non-self-conflicting operations, the performance of the point-only scenarios increased significantly. At the same time, the influence on the other scenarios was insufficient.

\end{document}